# DFT prediction of new o-MAX phases: Mo₂A₂AlC₃ (A = Zr, Nb, Ta) for next-generation thermal barrier coatings

M. I. A. Tanim, C. Talukder, S. S. Saif, Labib H. K. Adnan, N. Jahan, M. M. Hossain, M. M. Uddin, M. A. Ali*

Advanced Computational Materials Research Laboratory (ACMRL), Department of Physics

Chittagong University of Engineering and Technology (CUET), Chattogram-4349, Bangladesh

## Abstract

The incorporation of o-MAX phases, characterized by out-of-plane atomic arrangements, provides valuable extensions to the MAX phase family, driven by their superior thermo-mechanical properties, which are suitable for high-temperature applications. In this research, three novel o-MAX phases, Mo₂A₂AlC₃ (A = Zr, Nb, Ta), have been newly explored, and their structural geometry, electronic properties, mechanical behavior, thermodynamic characters, and optical response have been comprehensively investigated employing density functional theory (DFT) for the first time. The formation energy ($E_F$), phonon dispersion curve (PDC), and elastic constants ($C_{ij}$) were used to validate the stability of the predicted compounds. Study of the band structure, density of states (DOS), and charge density mapping revealed the metallic nature and presence of a mixed bonding within Mo₂A₂AlC₃ (A = Zr, Nb, Ta) phases. Mechanical properties were thoroughly evaluated through calculations of elastic constants, elastic moduli, fracture toughness, machinability, and hardness. Variations in mechanical properties were explained via Mulliken population analysis and charge density mapping. Poisson's ratio, Pugh ratio, and Cauchy pressure values indicate a brittle nature for these materials. Additionally, elastic anisotropy was examined through 2D and 3D diagrams of stiffness moduli and established anisotropy indicators. The Debye temperature ($\Theta_D$), Grüneisen parameter, lattice thermal conductivity ($k_{ph}$), minimum lattice thermal conductivity ($K_{min}$), melting point ($T_m$), and thermal expansion coefficient ($TEC$) were studied, making these compounds promising candidates for next-generation thermal barrier coating (TBC) applications. The optical properties were thoroughly analyzed, with reflectivity spectra (exceeding 44%) implies that these materials can efficiently reduce solar heating across various energy ranges.

Corresponding Author: ashrafphy31@cuet.ac.bd

## 1. Introduction

Emerging progressions in MAX phase research have intensified their influence across the domain of materials science due to their remarkable strength under stress and high-temperature stability while simultaneously demonstrating both metal-like conductivity and ceramic-like durability. They include a wide range of ternary compounds, such as carbides, nitrides, and borides, expressed by the generic stoichiometric expression $M_{n+1}AX_n$, where M is an early transition metal, A is from group IIIA or IVA, and X is C, N, or B, with n equal to 1, 2, or 3 [1]. Variation in the integer value (n = 1, 2, 3) governs the classification of MAX phases into the respective categories: $M_2AX$ (211), $M_3AX_2$ (312), and $M_4AX_3$ (413). Barsoum et al. [1] [2] first studied carbides and nitrides in the 1990s, which led to their widespread recognition, although Nowotny et al. [3] [4] laid the groundwork for MAX phases in the 1960s. The metallic nature of MAX materials comes from the alternating metallic A-layers, while their ceramic qualities arise from the MX layers [1]. These materials offer improved machinability, enhanced electrical and thermal conductivity, and excellent thermal shock resistance, comparable to metals and alloys. Conversely, they also have high melting or decomposition points and strong elastic stiffness, like ceramics [2]. Their unique combination of metallic and ceramic features enables various applications, such as high-temperature coatings for nuclear accident-tolerant fuel (ATF), concentrated solar power (CSP), catalysis, and as precursors for MXenes [1] [2] [3].

Recent studies have extensively forecasted and synthesized thermodynamically stable MAX phase compounds, leading to the discovery of out-of-plane ordered subgroups known as o-MAX phases. These are outlined within the typical stoichiometric expression $M'_2M''_nAX_{n+1}$, where n is 1 or 2, reflecting different atomic layer arrangements [2]. In o-MAX phases, the outer layer is composed of M' atoms, with A atoms positioned at the surface. Unlike the A-atoms, M'' atoms are situated within an internal layer surrounded by X-atoms, resulting in structured atomic stacking that enhances material properties [2] [4]. Liu et al. [5] discovered the first o-MAX phase by alloying M elements in $Cr_2TiAlC_2$. This result promptly led to the synthesis of $Mo_2ScAlC_2$ [6], $Mo_2TiAlC_2$ [7] [8], $Mo_2Ti_2AlC_3$ [7], and $Mo_2Nb_2AlC_3$ [9]. The investigation revealed that $Mo_2TiAlC_2$ and $Mo_2Ti_2AlC_3$ compounds possess excellent mechanical performance, making them suitable for structural applications. Compared to conventional MAX phases, the o-MAX phases display superior mechanical behavior due to their atomic configuration. As a result, their validity for structural ceramic applications highlights the o-MAX family's role in materials designed to withstand diverse stress and heat. A broad range of studies have been conducted on Mo-based MAX phases, revealing their practical applications through crystallographic structure and operational features. $Mo_2TiAlC_2$ has been thoroughly examined by Hadi et al. [10], who provided a rigorous assessment of its structural, mechanical, and electronic properties, as well as estimated its Vickers hardness and Debye temperature. The study offers an inclusive overview of the material's thermo-mechanical response, confirming its potential for structural applications in demanding conditions. Using DFT, Hadi et al. [11] also investigated $Mo_2ScAlC_2$, specifically examining its mechanical properties, chemical bonding nature, and defect mechanisms. The thermodynamic, optical, and elastic features of $Mo_2TiAlC_2$ were anticipated by Gao et al. within the pressure interval of 0 to 100 GPa [12]. Wyatt et al. [9] examined $Mo_2Nb_2C_3T_x$ to assess its atomic-level structure and effectiveness in hydrogen-based catalysis. The compounds showed substantial promise in sustainable energy applications due to their enhanced catalytic activity. B. Anasori et al. [7] provided valuable insights into the synthesis, structural properties, and phase stability of $Mo_2TiAlC_2$ and $Mo_2Ti_2AlC_3$. Their synthesis involved controlled heating of a combination of elements, illustrating the diverse functionality and flexibility of Mo-based MAX phases. Using solid-state reaction, Wen et al. [13] reported the synthesis of $Mo_2VAlC_2$ and $Mo_2V_2AlC_3$ o-MAX phases with high-grade purity, which subsequently served as precursors for o-MXene production, where HF solution etching facilitated the formation of $Mo_2VC_2T_x$ and $Mo_2V_2C_3T_x$, processed in both thin film and powdered states. Corresponding to this experimental pathway, Mia et al. [14] provided valuable insights into the physical properties of $Mo_2VAlC_2$ and $Mo_2V_2AlC_3$, performing in-depth first-principles analyses. Recently, several studies have revealed substantial progress in Mo-based o-MAX phases. We have been motivated by the growing attention to o-MAX phases, owing to their exceptional intrinsic properties and promising applications in advanced technologies. Consequently, in this study, we have selected the o-MAX phases $Mo_2A_2AlC_3$ (A = Zr, Nb, Ta), which are being predicted for the first time, and explored their properties for potential applications.

Currently, there are no theoretical or experimental studies on $Mo_2A_2AlC_3$ (where A = Zr, Nb, Ta) MAX phases, despite significant advancements in MAX-phase research. To address this gap, we conducted comprehensive DFT calculations to analyze the structural, electronic, mechanical, thermodynamic, and optical properties of these materials. The primary goal of this study is to explore new o-MAX phases, while the secondary objective is to assess their suitability for extreme thermal conditions, particularly for use as thermal barrier coatings (TBC). Our results suggest that these materials have potential applications in TBC technologies.

## 2. Computational methodology

A quantum mechanical modeling was carried out utilizing the CASTEP code [15], which incorporates the plane-wave pseudo-potential approximation within the density functional theory (DFT) framework [16]. The electron-ion interactions and exchange-correlation functionals were evaluated employing the generalized gradient approximation (GGA) - Perdew–Burke–Ernzerhof (PBE) [17] and its solid-state optimized counterpart, PBEsol [18]. For DFT calculations, the PBE has been used widely, but PBEsol offers enhanced precision in estimating lattice parameters for crystalline materials [19]. Accurate measurements of lattice constants play a crucial role in refining computed values, with a significant impact on factors that determine the mechanical response. Accordingly, in this investigation, PBEsol was employed alongside the commonly adopted GGA-PBE scheme to ensure computational reliability [17] [18]. The ultrasoft pseudopotential of the Vanderbilt type is used to represent electron-ion interactions and the electron configurations of C - $2s^2 \, 2p^2$, Al - $3s^2 \, 3p^1$, Zr - $4s^2 \, 4p^6 \, 4d^2 \, 5s^2$, Nb - $4s^2 \, 4p^6 \, 4d^4 \, 5s^1$, Ta - $5d^3 \, 6s^2$, and Mo – $4s^2 \, 4p6 \, 4d^5 \, 5s^1$ orbitals. To ensure convergence of the plane wave expansion, the cutoff energy was set to 650 eV. The first Brillouin zone was integrated using an $8 \times 8 \times 3$ k-point grid constructed via the Monkhorst–Pack approach [20]. Relaxation of the structures was carried out using the Broyden-Fletcher-Goldfarb-Shanno (BFGS) technique [21], while the electronic structures are computed employing a density mixing scheme. Structural optimization was achieved by the following tolerance criteria: $5 \times 10^{-6}$ eV/atom for total energy, 0.01 eV/Å for atomic forces, $5 \times 10^{-4}$ Å for uppermost ionic displacement, and 0.02 GPa for the maximum stress. Elastic properties were evaluated utilizing the finite "stress-strain" method within the framework of DFT. The vibrational characteristics, including phonon dispersion and density of states, were determined by employing density functional perturbation theory (DFPT) based on a finite displacement approach. During these calculations, a higher k-point sampling of $9 \times 9 \times 2$ and a consistent cutoff energy of 550 eV were applied across the Brillouin zone.

## 3. Results and discussion
### 3.1 Structural properties:

The crystal structure of the o-MAX compounds $Mo_2A_2AlC_3$ (A = Zr, Nb, Ta), which belong to the hexagonal lattice with the $P6_3/mmc$ space group (No. 194), is illustrated in Fig. 1. As depicted in Fig. 1, two individual formula units reside in the unit cell, summing up to 16 atoms overall.

In the optimized structure (Fig. 1), Mo, A, and Al atoms occupy the (4e), (4f), and (2c) Wyckoff sites with fractional coordinates (0, 0, $Z_M$), (1/3, 2/3, 0.058108), and (1/3, 2/3, 0.25), respectively. Conversely, two C atoms are located at (4f) (2/3, 1/3, 0.115787) and (2a) (0, 0, 0). To find the most stable configurations, the relaxation process of the structure $Mo_2A_2AlC_3$ (A = Zr, Nb, Ta) proceeded until the system achieved its lowest total energy. The data in Table 1 present the lattice parameters ($a$, $c$), the hexagonal ratio ($c/a$), internal parameters ($Z_M$), density ($\rho$), volume ($v$), formation energy ($E_F$), and summarize available experimental results for comparison. The agreement between the obtained results and their respective reference values demonstrates the accuracy of the applied computational approach [15]. Hug's distortion indices (DIs) [22] were used to assess structural distortion using the ($c/a$) ratio and the $Z_M$ parameter. The following formulas can be employed to estimate this distortion corresponding to the octahedron site [$M_6X$] and the trigonal prism site [$M_6A$] [23].

$$O_r = \frac{\sqrt{3}}{2\sqrt{4Z_m^2\left(\frac{c}{a}\right)^2 + \frac{1}{12}}} \qquad (1)$$

$$P_r = \frac{1}{\sqrt{\frac{1}{3} + \left(\frac{1}{4} - Z_m\right)^2\left(\frac{c}{a}\right)^2}} \qquad (2)$$

When the structure is free from distortion, the octahedron and trigonal metrics are expected to be identical, with $O_r$ and $P_r$ values equal to 1 [23]. The discrepancy from 1 is used for assessing the polyhedron's distortion, with minimal distortion implying enhanced structural stability [23]. According to Table 1, the octahedral site shows a higher degree of distortion compared to the prismatic site, as the values of $O_r$ are significantly less than 1, while $P_r$ is close to 1. There have also been reports of such structural distortions in other MAX phase components [18].

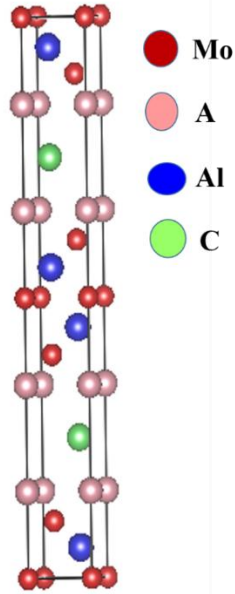

**Fig. 1**: The schematic unit cell of $Mo_2A_2AlC_3$ (A = Zr, Nb Ta).

**Table 1**: The optimized lattice constants ($a$ and $c$ in Å), $c/a$ ratio, internal parameters ($Z_M$ in Å), volume ($V$ in Å³), density ($\rho$ in g/cm³), distortion parameters of octahedral ($O_r$)and trigonal prisms ($P_r$) and formation energy ($E_F$ in eV) of $Mo_2A_2AlC_3$ (A = Zr, Nb, Ta).

| Phases | $a$(Å) | $c$(Å) | $c/a$ | $Z_M$ | $V$ (Å³) | $\rho$(g/cm³) | $O_r$ | $P_r$ | $O_r/P_r$ | $E_F$ | Ref. |
|---|---|---|---|---|---|---|---|---|---|---|---|
| $Mo_2Zr_2AlC_3$ | 3.1502 | 23.999 | 7.618 | 0.163 | [a]206.267 | 7.04 | 0.346 | 1.138 | 0.304 | -0.506 | [a]This |
| | 3.1750 | 24.105 | 7.592 | | [b]210.451 | 6.90 | 0.347 | 1.140 | 0.304 | | [b]This |
| $Mo_2Nb_2AlC_3$ | 3.1026 | 23.551 | 7.590 | 0.156 | 196.341 | 7.45 | 0.362 | 1.090 | 0.332 | -0.349 | [a]This |
| | 3.1026 | 23.551 | 7.590 | | 196.341 | 7.45 | 0.362 | 1.090 | 0.332 | | [b]This |
| $Mo_2Ta_2AlC_3$ | 3.1436 | 23.797 | 7.569 | 0.158 | 203.665 | 10.05 | 0.359 | 1.105 | 0.325 | -0.386 | [a]This |
| | 3.1559 | 23.877 | 7.565 | | 205.953 | 9.94 | 0.359 | 1.105 | 0.325 | | [b]This |
| | [*]2.977 | [*]23.085 | [*]7.753 | [*]0.156 | [*]177.297 | [*]6.68 | [*]0.360 | [*]1.050 | [*]0.342 | [*]-0.47 | [13] |
| [a] computed value using GGA PBEsol [18], [b] calculated values using GGA-PBE [17]. [*] Reference [13]. | | | | | | | | | | | |

## 3.2 Stability

For several reasons, assessing the stability of a compound is crucial, as it provides critical insights into its optimal synthesis parameters. It also enables a comprehensive evaluation of its durability across diverse environments, including thermal, compressive, and mechanical pressures. This work presents a comprehensive computational investigation encompassing (i) chemical stability by computing the formation energy, (ii) dynamical stability via phonon dispersion analysis, and (iii) mechanical stability by computing the elastic constants ($C_{ij}$).

Evaluation of the formation energy for the analyzed compounds was carried out using the following expression [24]:

$$E_{for}^{Mo_2A_2AlC_3} = \frac{E_{total}^{Mo_2A_2AlC_3} - \left(lE_{solid}^{Mo} + mE_{solid}^{A} + nE_{solid}^{Al} + pE_{solid}^{C}\right)}{l + m + n + p} \qquad (3)$$

Here, $E_{total}^{Mo_2A_2AlC_3}$ denotes the compound's total energy following structural relaxation of the unit cell. The energies of the individual components Mo, A, Al, and C are indicated by $E_{solid}^{Mo}$, $E_{solid}^{A}$, $E_{solid}^{Al}$, and $E_{solid}^{C}$, respectively. Within the unit cell, the number of Mo, A, Al, and C atoms is represented by *l, m, n,* and *p* sequentially. The computed formation energy ($E_F$) for Mo₂Zr₂AlC₃, Mo₂Nb₂AlC₃, and Mo₂Ta₂AlC₃ are -0.506 eV/atom, -0.349 eV/atom, and -0.386 eV/atom, in sequence. Herein, the compounds exhibit chemical stability, as evidenced by their negative $E_F$ values. The order of chemical stability is as follows: Mo₂Zr₂AlC₃ > Mo₂Ta₂AlC₃ > Mo₂Nb₂AlC₃, suggesting that Mo₂Zr₂AlC₃ possesses superior structural robustness compared to the other analyzed compounds.

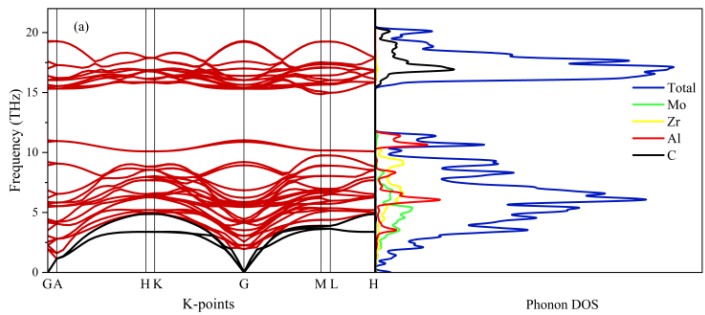

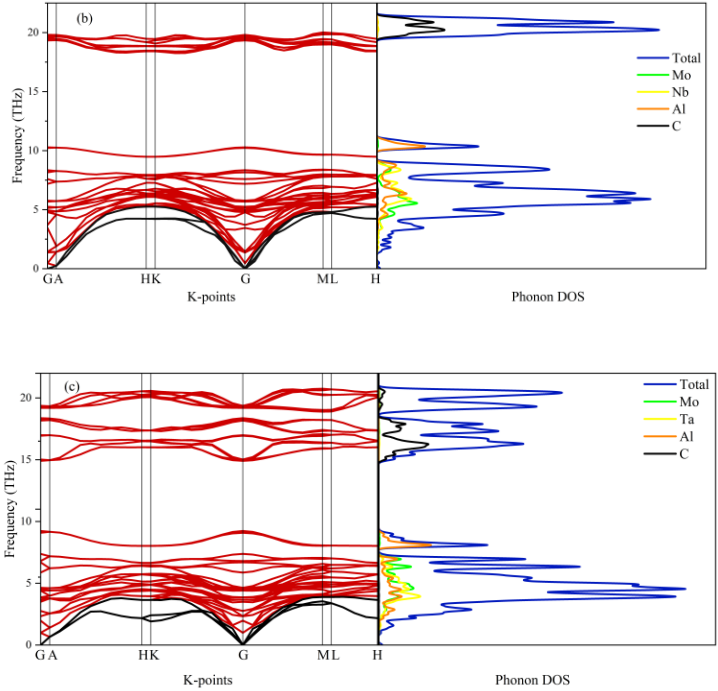

**Fig. 2**: The phonon dispersion curves (PDCs) and phonon density of states (PHDOS) of $Mo_2A_2AlC_3$ (A = Zr, Nb, Ta) (a, b, c), respectively.

A material's dynamic stability is one of the key factors that influences its applicability when subjected to time-varying mechanical loads. Computations of phonon dispersion as well as phonon density of states were carried out for $Mo_2A_2AlC_3$ (A = Zr, Nb, Ta) phases to verify the dynamic equilibrium alongside the influence of vibrations on thermodynamic attributes. Figs. 2(a), 2(b), and 2(c) illustrate the PDC and PHDOS of $Mo_2A_2AlC_3$ (A = Zr, Nb, Ta) compounds following the principal symmetry directions within the Brillouin zone of the crystal structure. The material's stability was investigated through phonon frequency evaluations spanning the entire Brillouin zone. None of the compounds has negative phonon frequencies at any $k$-points, as visualized in Fig. 2(a, b, c), which indicates that the phases remain dynamically stable when subjected to time-varying mechanical stresses at room temperature. The unit cell of the examined o-MAX phases contains 16 atoms, which generate 48 vibrational modes in the phonon dispersion curves; among these, three correspond to acoustic modes, while the remaining 45 are optical modes. The acoustic branches are the lowest branches on the dispersion curve. When $k$ is small, the acoustic dispersion that describes the correlation between frequency and wave vector has the form of $\omega = vk$ [25].

These phonons are produced when atoms in a crystal vibrate collectively, deviating from their symmetry locations.

A crystal's higher branches, known as optical modes, mainly determine its optical properties. Due to photon-induced excitation, the atoms oscillate out of phase, generating optical phonons. The lower optical branches and acoustic modes overlap, preventing the formation of a gap in vibrational frequencies between the two modes. For improved visualization of the vibrational spectrum, the PHDOS of $Mo_2A_2AlC_3$ phases is illustrated adjacent to the phonon dispersion diagrams. Whereas the faint peaks are caused by non-flat bands, and the prominent peaks correspond to the PDCs' flat modes. In $Mo_2A_2AlC_3$ (A = Zr, Nb, Ta), the vibrational modes within the (3–10 THz) range are predominantly associated with the vibrations of (Mo, Zr, Al) atoms. In contrast, the frequency beyond 15 THz is dominated by contributions from the C atoms, as depicted in Figs. (a, b, c).

### 3.3 Electronic properties

### 3.3.1 Electronic band structure and density of states (DOS)

To examine the electronic conductivity, electron mobility, and transport behavior of $Mo_2A_2AlC_3$ (A = Zr, Nb, Ta) MAX phases, their electronic band structure is computed along standard symmetry directions in reciprocal space based on optimized unit cell constants. Within the band structure in Fig. 3, the Fermi level ($E_F$) is indicated by a green horizontal line at the zero-energy scale. The band structure exhibits a significant amount of overlaps between valence-conduction bands near the Fermi energy, confirming the metallic behavior of the material, similar to other MAX Phases [26] [27] [28] [29]. The depicted compounds illustrated in Fig. 3 (a–c) exhibit anisotropic electronic conductivity due to band energy variation between the basal plane (*ab*) and the *c*-axis. Band dispersion in the Brillouin zone occurs along *A–H*, *K–G*, *G–M*, and *L–H* in the basal (*ab)* plane, while *G–A*, *H–K*, and *M–L* paths represent c-axis dispersion. This anisotropy in band dispersion correlates with reduced electronic mobility along the c-axis compared with the *ab*-plane. This behavior arises from directional differences in the electron effective-mass tensor across the crystal [30]. As a result, anisotropy in the effective mass tensor, with greater values along the c-axis, reflects a potential decline in electronic conductivity in that direction [31]. MAX phases exhibit a layered configuration in which atomic bonding is stronger and more delocalized within the basal planes than along the *c*-axis. Cubic materials exhibit an isotropic nature owing to the same atomic arrangement along all axes. Conversely, hexagonal or layered structured materials exhibit an anisotropic nature, as their atomic configurations vary along different directions [32].

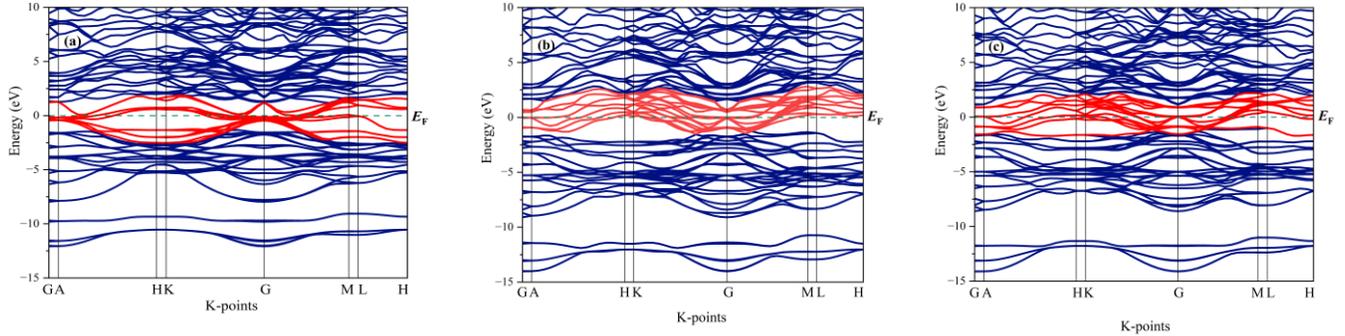

**Fig. 3:** Band structure of (a) Mo$_2$Zr$_2$AlC$_3$, (b) Mo$_2$Nb$_2$AlC$_3$, (c) Mo$_2$Ta$_2$AlC$_3$.

To explore the underlying electronic transport behavior and atomic bonding nature of Mo$_2$A$_2$AlC$_3$ (A = Zr, Nb, Ta), the total and partial density of states (TDOS and PDOS) have been calculated, which delineate orbital hybridization and individual atomic orbital contributions near the Fermi level. Analysis of the density of states indicates that low-energy bonding states originate mainly from the hybridization of Mo-4$p$, 4$d$, Zr-4$d$, Nb-4$d$, and Ta-4$d$ orbitals, as well as the $p$ orbitals of Al and C. This pronounced Mo-C covalency is responsible for the structural stability of the MAX phases. The negligible hybridization of Mo-5$s$, Zr-5s, Nb-5$s$, 4$p$, Ta-6$s$, and Al-3$s$ orbitals below the Fermi level with C-2$s$ orbitals reflects the major non-bonding nature. In Mo$_2$Nb$_2$AlC$_3$, anti-bonding states remain unoccupied and weakened due to strong Mo-C covalent hybridization, enhancing structural stability. Mo$_2$Ta$_2$AlC$_3$ exhibits partially filled anti-bonding orbitals primarily contributed by Mo-4$d$ and Ta-5$d$ orbitals, which leads to an equal balance of strong Mo-Ta ($d$-$d$) bonding. At the same time, Mo$_2$Zr$_2$AlC$_3$ shows a decline in occupancy of Zr-$d$ states, offering promising electronic characteristics for use in advanced technologies. Additionally, all three compounds have partially filled states, where the pseudo-gap is located slightly right above the Fermi level, indicative of a metallic nature driven by increased electron mobility. The TDOS and PDOS spectra of Mo$_2$A$_2$AlC$_3$ are displayed in Fig. 4 (a–c). The lower valence band near E$_f$ exhibits peaks arising from the hybridization between Mo-4$d$ 4$p$, Zr-4d/Nb-4$d$/Ta-5$d$, Al-3$p$, and C-2$p$ states, where the Mo-4$d$ and Al-3$p$ contribute most significantly. The conduction band, located above the Fermi level, is predominantly derived from the Mo-4$d$, Zr-4d/Nb-4$d$/Ta-5$d$, Al-3s, and C-2$p$ states. As depicted in Fig. 5, the TDOS of Mo$_2$A$_2$AlC$_3$ (A = Zr, Nb, Ta), which exhibit a non-zero, finite density at the Fermi level, indicates their inherent metallicity. Additionally, a slight shift of the energy peaks (highlighted by red lines) toward lower energies relative to the Fermi level is observed. The red lines mark the peak positions that arise from orbital hybridization among contributing electronic states, leading to the initiation of stable covalent bonding between the atoms. There is a direct relationship between DOS peak energies and bond strength; reduced peak energies denote stronger covalent interactions. Accordingly, the observed peak positions in $Mo_2Nb_2AlC_3$ suggest enhanced bonding among its constituent atomic states. Hence, the prominent characteristics in the TDOS of $Mo_2A_2AlC_3$ (A = Zr, Nb, Ta) suggest a stronger bond between the atomic states. The order of bonding strength is found to be: $Mo_2Nb_2AlC_3 > Mo_2Zr_2AlC_3 > Mo_2Ta_2AlC_3$.

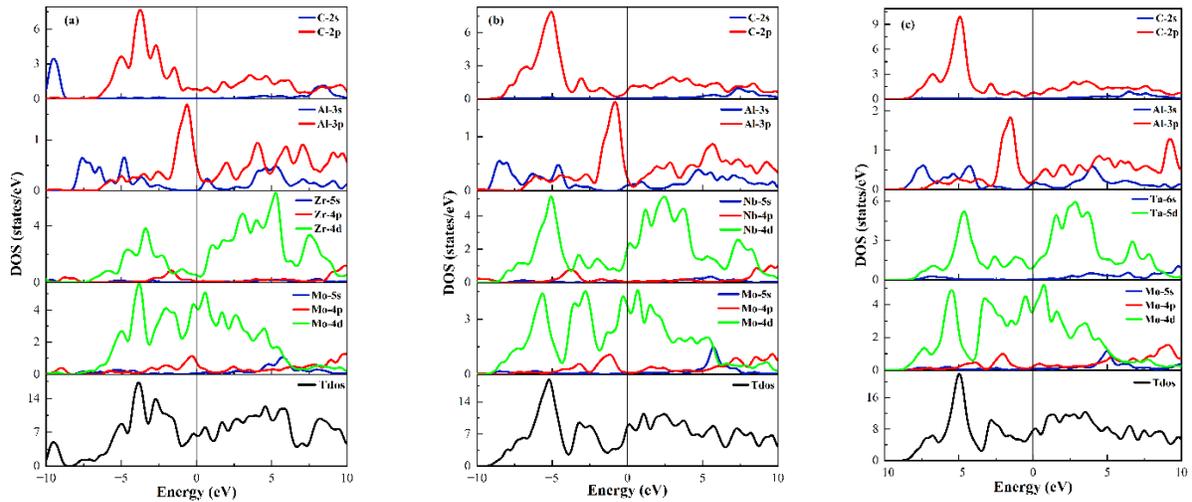

**Fig. 4:** TDOS and PDOS of (a) $Mo_2Zr_2AlC_3$, (b) $Mo_2Nb_2AlC_3$, (c) $Mo_2Ta_2AlC_3$.

Electron localization, chemical bonding characteristics, and the nature of atom-to-atom interactions can be visualized through charge density mapping (CDM), which is depicted along the (110) plane and expressed in units of $e/Å^3$. It helps to identify the accumulation of negative charge (high-density region) and positive charge (low-density region), where regions shaded in blue indicate high electron density, in contrast to red areas that reflect low electron density. A high accumulation of electron charge forms between two atoms, resulting in a covalent bond, as determined by the difference in electronegativity. In contrast, ionic bonding originates from the coulombic force between positively and negatively charged ions, with one atom having a net negative charge due to the accumulation of electrons, and the other becomes positively charged by losing electrons. As depicted in Figs. 5(a), (b), and (c), a high charge concentration is observed in the regions between the C sites in $Mo_2Zr_2AlC_3$, $Mo_2Nb_2AlC_3$, and $Mo_2Ta_2AlC_3$, resulting in a strong covalent Mo-C (M-X) bond.

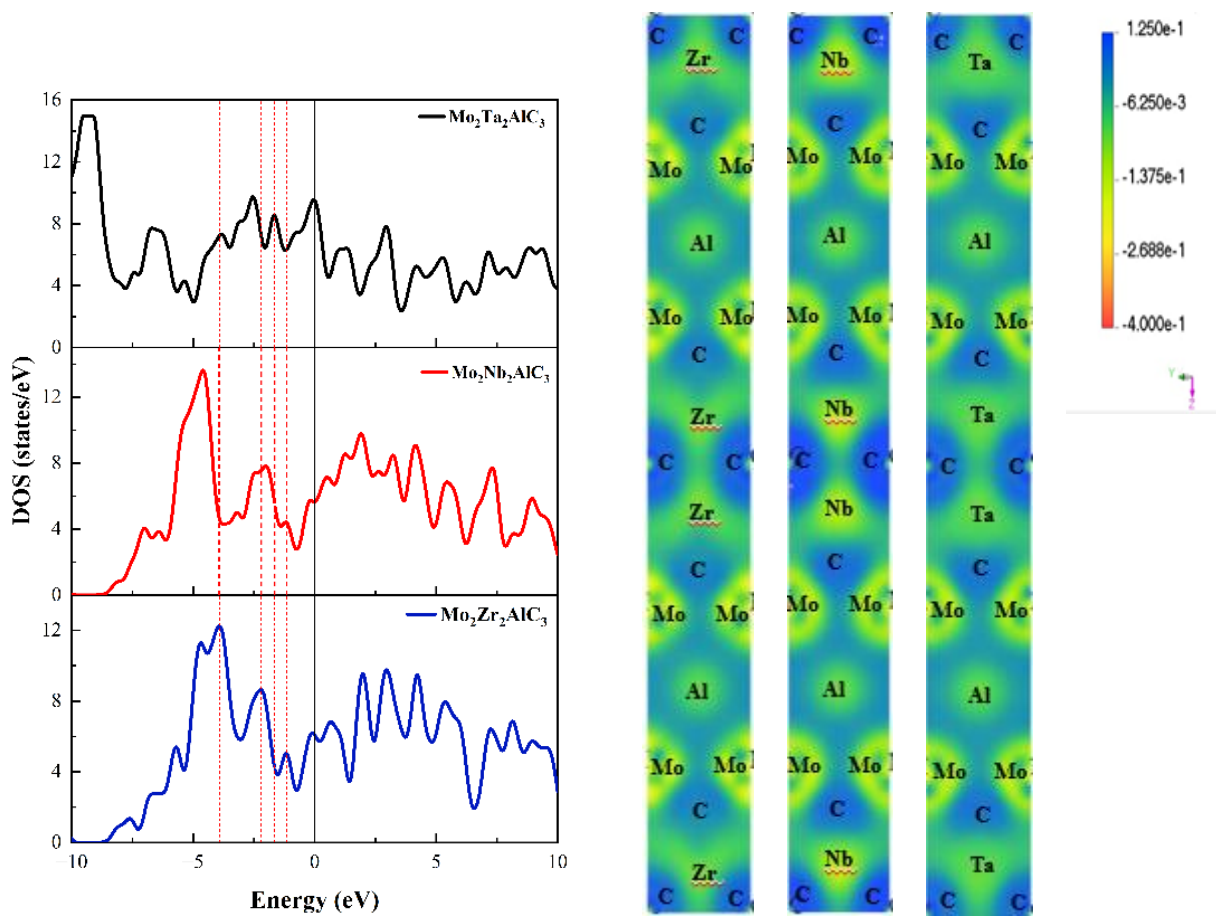

Fig. 5: The total DOS and charge density mapping of $Mo_2A_2AlC_3$ (A = Zr, Nb, Ta).

As Al is less electronegative than Mo, Fig. 5 reveals minimal electron sharing between Mo and Al, indicating a metallic bonding nature due to the low electron density in the bonding region. According to the analyses of DOS and CDM, the $Mo_2Nb_2AlC_3$ compound is anticipated to exhibit the highest values of stiffness constant, elastic moduli, and hardness parameter.

### 3.3.2 Mulliken populations and bond population analysis

Mulliken population analysis provides a quantitative method for assessing charge distribution and interatomic bonding properties, particularly those influenced by the participation of transition metal d-orbitals. Sánchez-Portal et al. introduced a formula that maps delocalized electronic states into localized atomic orbitals for detailed analysis [33] and then applies the Mulliken method for population analysis. Table 2 presents the Mulliken atomic charge and EVC (effective valence charge), which define whether a chemical bond is ionic or covalent. Both the covalent and ionic natures are reflected in MAX phase compounds when they have a non-positive effective valence charge value (EVC). As the value approaches zero, the bond exhibits mainly an ionic nature. A value of zero represents a perfectly ionic bond. A zero value represents perfectly ionic bonding. On the other hand, an increase from zero to a positive value is reflected in elevated covalent bonding. According to Table 2, the transition metals Mo, Nb, Zr, and Ta have positive Mulliken charges, while C has negative values. This indicates that C atoms acquire electrons during bonding, whereas Mo, Nb, Zr, and Ta atoms lose electrons. Mulliken population analysis reveals that 0.64 |e|, 0.76 |e|, 0.64 |e|, 0.70 |e|, 0.67 |e|, 0.74 |e| are transmitted from M (Mo, Nb, Zr, Ta) atoms to the C atoms. As a result, during the bonding process, C becomes an electron acceptor, whereas Mo, Nb, Zr, and Ta function as electron donors. As a result, an ionic bond is formed among these compounds.

Another important parameter that provides a quantitative measure of bonding and anti-bonding strengths within the chemical bond is the Bond overlap population (BOP) [34] [35]. When the electronic population of the two atoms fails to interact substantially, BOP is found to be zero. A positive BOP value indicates a bonding nature between neighboring atoms, whereas a negative value indicates an antibonding nature. A shorter bond length is generally associated with a higher bond overlap population (BOP), indicating stronger covalent interactions between the atoms. As illustrated in Table 3, the Mo-C and Nb-C bonds in $Mo_2Nb_2AlC_3$ exhibit shorter bond lengths (2.029 Å & 2.151 Å) compared to those in analogous compounds, indicating a stronger covalent interaction and enhanced bond strength in $Mo_2Nb_2AlC_3$. The absence of negative bond overlap population (BOP) values throughout the structure suggests that no significant anti-bonding interactions are present between the essential atoms, thereby affirming the compound's stability in terms of bonding and electronic coherence. The bond ionicity trend can be reflected in the bond overlap populations. The bond ionicity trend can be reflected in the bond overlap populations. A bond's ionicity can be computed using the bond overlap population as follows: [36]

$$P_i = 1 - \exp\left[-|P_c - P^\mu| / P^\mu\right] \tag{4}$$

Here, $P^\mu$ represents the overlap population of a $\mu$-type bond, while $P_c$ denotes the overlap population of the bond in a purely covalent crystal. A covalent bond is represented by a value of 0 on the ionicity index, whereas a pure ionic bond is represented by a value of 1. A bond overlap population of $P_c = 2$ is taken in this context as reflective of a fully covalent interaction. The bond's level of metallicity is determined by the following expression [37] [38]

$$f_m = P^{\mu'}/P^\mu \tag{5}$$

Finally, the metallicity analysis reveals that in $Mo_2Zr_2AlC_3$, the C-Mo bond is the most metallic (0.189); in $Mo_2Nb_2AlC_3$, the C-Nb(1) bond shows the highest metallicity (0.03); and in $Mo_2Ta_2AlC_3$, the C-Mo bond is the most metallic (0.059), indicating variations in bond character depending on the transition metal.

**Table 2**: Mulliken atomic populations and Effective valence charge of $Mo_2A_2AlC_3$ (A = Zr, Nb, Ta) compounds.

| Compounds | atoms | $s$ | $p$ | $d$ | $f$ | Total | Charge (e) | Effective Valence charge EVC (e) |
|---|---|---|---|---|---|---|---|---|
| $Mo_2Zr_2AlC_3$ | C | 1.47 | 3.17 | 0.00 | 0.00 | 4.64 | -0.64 | - |
| | C | 1.48 | 3.27 | 0.00 | 0.00 | 4.76 | -0.76 | - |
| | Al | 0.92 | 1.70 | 0.00 | 0.00 | 2.63 | 0.37 | 2.63 |
| | Zr | 2.12 | 6.36 | 2.68 | 0.00 | 11.16 | 0.84 | 3.16 |
| | Mo | 2.28 | 6.68 | 5.05 | 0.00 | 14.01 | 0.01 | - |
| $Mo_2Nb_2AlC_3$ | C | 1.43 | 3.22 | 0.00 | 0.00 | 4.64 | -0.64 | - |
| | C | 1.44 | 3.26 | 0.00 | 0.00 | 4.70 | -0.70 | - |
| | Al | 0.88 | 1.80 | 0.00 | 0.00 | 2.68 | 0.32 | 2.68 |
| | Nb | 2.10 | 6.30 | 3.87 | 0.00 | 12.26 | 0.74 | 4.26 |
| | Mo | 2.20 | 6.62 | 5.09 | 0.00 | 13.90 | 0.10 | 5.9 |
| $Mo_2Ta_2AlC_3$ | C | 1.46 | 3.21 | 0.00 | 0.00 | 4.67 | -0.67 | - |
| | C | 1.50 | 3.24 | 0.00 | 0.00 | 4.74 | -0.74 | - |
| | Al | 0.95 | 1.75 | 0.00 | 0.00 | 2.70 | 0.30 | 2.7 |
| | Mo | 2.18 | 6.33 | 5.04 | 0.00 | 13.55 | 0.45 | 5.55 |
| | Ta | 0.33 | 0.50 | 3.73 | 0.00 | 4.56 | 0.44 | 4.56 |

**Table 3:** Calculated Mulliken bond overlap population.

| Compounds | Bond | $n^\nu$ | $d^\nu$ (Å) | $P^\mu$ | $P^{\mu^r}$ | Bond covalency (%) | Bond ionicity (%) | Bond metallicity (%) |
|---|---|---|---|---|---|---|---|---|
| $Mo_2Zr_2AlC_3$ | C-Mo | 4 | 2.132 | 1.32 | 0.025 | 66 | 40 | 18.93 |
| | C-Zr1 | | 2.282 | 1.00 | 0.025 | 50 | 63.2 | 2.5 |
| | C-Zr2 | | 2.308 | 0.65 | 0.025 | 33 | 87.4 | 3.84 |
| $Mo_2Nb_2AlC_3$ | C-Mo | 4 | 2.029 | 1.01 | 0.025 | 51 | 62 | 2.47 |
| | C-Nb1 | | 2.151 | 0.83 | 0.025 | 42 | 76 | 3.0 |
| | C-Nb2 | | 2.165 | 0.91 | 0.025 | 46 | 70 | 2.74 |
| $Mo_2Ta_2AlC_3$ | C-Mo | 4 | 2.074 | 0.52 | 0.031 | 26 | 94 | 5.96 |
| | C-Ta1 | | 2.275 | 1.51 | 0.031 | 76 | 28 | 2.05 |
| | C-Ta2 | | 2.281 | 1.14 | 0.031 | 57 | 53 | 2.72 |

## 3.4 Mechanical properties

### 3.4.1 Stiffness constants and elastic moduli

Analyzing the mechanical behavior of crystalline materials provides insight into their response to various applied loads or forces. Additionally, it facilitates predicting their potential uses in various fields by assessing physical traits, including elastic parameters, stiffness, brittleness, resistance to deformation, fracture toughness, machinability, damage tolerance, and directionally variable elastic behavior. This study employed a stress–strain-based computational framework to investigate the aforementioned properties of novel o-MAX phase compounds, $Mo_2A_2AlC_3$ (A = Zr, Nb, Ta) [39] [40] [41]. Using the computed elastic parameters, the polycrystalline mechanical moduli were computed employing Hill's approximation, which averages the Voigt and Reuss boundary predictions [42]. Since the $Mo_2A_2AlC_3$ (A = Zr, Nb, Ta) phase under consideration has a hexagonal crystal structure, it possesses six stiffness constants ($C_{11}$, $C_{12}$, $C_{13}$, $C_{33}$, $C_{44} = C_{55}$, and $C_{66}$) wherein five are independent and $C_{66}$ is dependent as $C_{66} = (C_{11}\text{-}C_{12})/2$ [43]. The results for the computed elastic constants are summarized in Table 4, with their visual representation depicted in Fig. 6(a), for the examined o-MAX phase compounds, along with those of their respective reference phases, to facilitate a comparative evaluation.

**Table 4:** The stiffness constants, elastic moduli, Poisson's ratio ($v$), Cauchy Pressure ($CP$), $f$-index, Pugh's ratio ($G/B$), machinability index ($B/C_{44}$), fracture toughness ($K_{IC}$), brittleness index ($M_B$) and hardness parameters of $Mo_2A_2AlC_3$.

| Parameters | Functional | $Mo_2Zr_2AlC_3$ | $Mo_2Nb_2AlC_3$ | $Mo_2Ta_2AlC_3$ | *$Mo_2Ti_2AlC_3$ | *$Mo_2V_2AlC_3$ |
|---|---|---|---|---|---|---|
| $C_{11}$ (GPa) | GGA-PBEsol | 397 | 442 | 428 | 424 | 473 |
| | GGA-PBE | 393 | 416 | 416 | - | - |
| $C_{33}$ (GPa) | GGA-PBEsol | 346 | 370 | 328 | 382 | 440 |
| | GGA-PBE | 329 | 347 | 329 | - | - |
| $C_{44}$ (GPa) | GGA-PBEsol | 167 | 149 | 137 | 163 | 145 |
| | GGA-PBE | 159 | 114 | 128 | - | - |
| $C_{12}$ (GPa) | GGA-PBEsol | 114 | 137 | 133 | 130 | 147 |
| | GGA-PBE | 123 | 132 | 136 | - | - |
| $C_{13}$ (GPa) | GGA-PBEsol | 134 | 138 | 104 | 144 | 158 |
| | GGA-PBE | 129 | 126 | 116 | - | - |
| $C_{66}$ (GPa) | GGA-PBEsol | 141 | 152 | 147 | 147 | 163 |
| | GGA-PBE | 135 | 141 | 140 | - | - |
| $B$ (GPa) | GGA-PBEsol | 211 | 230 | 205 | 230 | 257 |
| | GGA-PBE | 208 | 215 | 210 | - | - |
| $G$ (GPA) | GGA-PBEsol | 143 | 145 | 140 | 149 | 152 |
| | GGA-PBE | 138 | 126 | 132 | - | - |
| $Y$ (GPa) | GGA-PBEsol | 352 | 361 | 342 | 367 | 380 |
| | GGA-PBE | 340 | 317 | 328 | - | - |
| $v$ | GGA-PBEsol | 0.22 | 0.23 | 0.22 | 0.23 | 0.25 |
| | GGA-PBE | 0.22 | 0.25 | 0.24 | - | - |
| $CP$ (GPa) | GGA-PBEsol | -52.45 | -11.87 | -3.92 | -33 | 2 |
| | GGA-PBE | -36.15 | 18.08 | 7.84 | - | - |
| $f$ | GGA-PBEsol | 1.147 | 1.308 | 1.574 | 1.118 | 1.078 |

| | GGA-PBE | 1.293 | 1.338 | 1.437 | - | - |
|---|---|---|---|---|---|---|
| $G/B$ | GGA-PBEsol | 0.67 | 0.63 | 0.68 | 0.64 | 0.59 |
| | GGA-PBE | 0.66 | 0.58 | 0.62 | - | - |
| $B/C_{44}$ | GGA-PBEsol | 1.26 | 1.54 | 1.49 | 1.41 | 1.77 |
| | GGA-PBE | 1.30 | 1.88 | 1.63 | - | - |
| $K_{IC}$ (MPa.m$^{1/2}$) | GGA-PBEsol | 2.67 | 2.78 | 2.59 | 2.77 | 2.94 |
| | GGA-PBE | 2.61 | 2.50 | 2.56 | - | - |
| $M_B$ (µm$^{-1/2}$) | GGA-PBEsol | 2.25 | 2.25 | 2.40 | 2.63 | 2.64 |
| | GGA-PBE | 2.30 | 2.51 | 2.43 | - | - |
| $H_{macro}$ (GPa) | GGA-PBEsol | 20.28 | 18.59 | 20.02 | 19.16 | 17.37 |
| | GGA-PBE | 19.21 | 15.17 | 17.28 | - | - |
| $H_{micro}$ (GPa) | GGA-PBEsol | 26.57 | 25.40 | 25.94 | 26.85 | 25.35 |
| | GGA-PBE | 25.14 | 20.67 | 22.92 | - | - |
| *Reference [7]. *Reference [14]. | | | | | | |

All the evaluated $C_{ij}$ values are found to be non-negative and meet the mechanical stability conditions initially suggested by Max Born [43] and subsequently refined by Mouhat et al: [44], $C_{11} > 0$, $C_{11}-C_{12} > 0$, $C_{44} > 0$, $C_{66} > 0$, $(C_{11} +C_{12})$ $C_{33}$- $2C_{13}{}^2 > 0$. Reported values in Table 4 comply with the stated criteria, demonstrating the mechanical integrity of the novel o-MAX compounds under investigation. These attributes position them as strong contenders for integration into emerging high-performance material technologies. Moreover, the elastic tensors provide valuable understanding of both the directional variations in elasticity and the underlying bonding mechanisms in solids. For instance, the fact that $C_{11} > C_{33}$ indicates a stronger atomic bonding along the [100] orientation compared to the [001] orientation. This means compressing the novel o-MAX phase Mo$_2$A$_2$AlC$_3$ (A = Zr, Nb, Ta) to the $a$-direction is tougher than compressing it to the $c$-direction. Moreover, $C_{11} > C_{33}$ reveals the compound's anisotropic nature. The ability of the material to withstand shear forces within the [100] orientation is evaluated by $C_{44}$. Since the magnitudes of $C_{11}$ and $C_{33}$ surpass that of $C_{44}$, shear deformation necessitates a lower impedance than unidirectional deformation for the o-MAX phases under consideration. In comparison to other single-crystal elastic constants, $C_{12}$ and $C_{13}$ are found to have comparatively lower values. Accordingly, the combined influence of these two factors induces strain along the $b$- and $c$-axes, while the corresponding stress is exerted along the *a*-axis. The moderate stiffness constants $C_{12}$ and $C_{13}$ suggest that if a substantial load is exerted along the *a*-axis, the compounds under consideration should be susceptible to shear along the *b*- and *c*-axes. The stiffness constant $C_{66}$, defined as [$C_{11}$-$C_{12}$/2], serves as an indicator of the shear rigidity along the [110] direction. Additionally, significant features of polycrystalline materials, such as bulk modulus (*B*), shear modulus (*G*), and Young modulus (*Y*) of the compounds of interest, are computed using the stiffness constants. Table 4 and Fig. 6(b) display the derived results. Where the bulk modulus was evaluated employing the Voigt [45] and Reuss [46] models in combination with Hill's average [47] expressed as: $B = (B_V + B_R)/2$. The Voigt component $B_V$ is defined as [2($C_{11} + C_{12}$) + $C_{33}$ + 4$C_{13}$]/9 and the Reuss component is derived from $B_R = C^2/M$; $C^2 = C_{11} + C_{12}C_{33} - 2C_{13}{}^2$; and $M = C_{11} + C_{12} + 2C_{33} - 4C_{13}$. $B_V$ represents the upper limit of the bulk modulus, while $B_R$ represents its lower limit, according to the Voigt and Reuss models. The modulus of rigidity, *G*, was derived as the arithmetic mean of Voigt ($G_V$) and Reuss ($G_R$): $G = (G_V + G_R)/2$. The expressions for $G_V$ and $G_R$, are $G_V = [M + 12C_{44} + 12C_{66}]/30$ and $G_R = (5/2) [C^2C_{44}C_{66}] / [3B_V C_{44} C_{66} + C^2 (C_{44} + C_{66})]$, respectively, with $C_{66} = (C_{11} - C_{12})/2$. A material's opposition to volumetric compression and the integrity of its atomic interactions are effectively gauged by the bulk modulus *B*. The high bulk modulus in $Mo_2A_2AlC_3$ (A =Zr, Nb, Ta) reveals robust chemical bonds and significant resistance to volume deformation. One of the fundamental parameters reflecting a material's ability to withstand transverse or plastic deformation is characterized by the shear modulus *G,* which also denotes its average shear ability. Owing to the high value of *G,* the novel o-Max phase $Mo_2A_2AlC_3$ (A = Zr, Nb, Ta) is expected to resist shear deformation during mechanical processing. A material's elastic response to directional tensile or compressive loads is characterized by Young's modulus *Y,* obtained via the subsequent formula: $Y = 9BG / (3B + G)$.

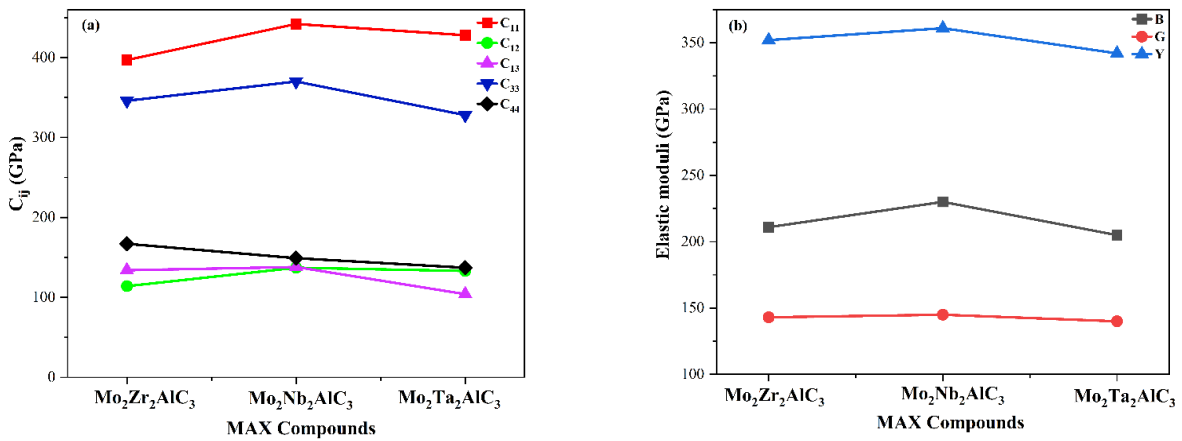

Fig. 6: (a) Elastic constants and (b) Elastic moduli of $Mo_2A_2AlC_3$ (A = Zr, Nb, Ta).

Table 4 and Fig. 6(b) illustrate that, in contrast to the other compounds under consideration, $Mo_2Nb_2AlC_3$ has the largest $Y$ value, representing its stronger stiffness. Among the investigated compounds, $Mo_2Nb_2AlC_3$ demonstrates superior stiffness, reflected by its elevated Young's modulus, implying improved resistance to external mechanical loads and deformation. A higher Young's modulus signifies that $Mo_2Nb_2AlC_3$ provides improved mechanical reliability and reduced deformability than other compounds being studied. This makes it suitable for high-performance machinery applications or aircraft parts that require high structural integrity. Moreover, thermal shock resistance ($R$) and Young's modulus ($Y$) have an inverse relationship: ($R \propto 1/E$) [47]. A reduced Young's modulus reflects enhanced impedance to thermal shock. For thermal barrier coatings (TBCs), materials with lower $Y$ values are therefore more appropriate. Table 4 and Fig. 6(b) demonstrate that $Mo_2Ta_2AlC_3$ has the lowest $Y$, making it an appealing option for TBC applications. It is pertinent to note that all assessed mechanical parameters align well with the DOS, BOP, and CDM [Fig. 5 and Table 3] outcomes and are expected to be consistent with Vickers' hardness findings, thereby confirming the reliability of the computational framework adopted in this study.

The resistance of a material to the growth of cracks can be quantified through its fracture toughness, $K_{IC}$. We have employed the following equation to estimate the fracture toughness: $K_{IC} = V_0^{\frac{1}{6}}.G.\left(\frac{B}{G}\right)^{\frac{1}{2}}$ [48]. Where $V_0$ stands for the measured volume, while $B$ and $G$ represent the bulk and shear modulus, respectively. Table 4 depicts high fracture toughness values (2.67 Mpam$^{1/2}$, 2.78 Mpam$^{1/2}$, 2.59 Mpam$^{1/2}$) for $Mo_2A_2AlC_3$ (A = Zr, Nb, Ta), respectively, indicating their ability to withstand the propagation of cracks. $Mo_2Nb_2AlC_3$ has a comparatively higher $K_{IC}$ value, implying that it can withstand fracture propagation better and be implemented in load-bearing circumstances, as it possesses the ability to tolerate increased pressure before cracking To gain insight into the MAX phases' capacity for damage tolerance, we defined the brittleness index $M_B$ as follows [49]: $M_B = \frac{H_V}{K_{IC}}$. Here, $H_V$ indicates Vickers hardness. Usually, lower $M_B$ values reflect better damage tolerance. It is evident from Table 4 that $Mo_2Zr_2AlC_3$ and $Mo_2Nb_2AlC_3$ possess relatively higher damage tolerance than $Mo_2Ta_2AlC_3$. The "$f$-index" is employed as an additional metric to characterize how uniformly atoms are bonded along the $a$- and $c$-axes. Table 4 also displays the $f$-value, derived through the subsequent formula [50]: $f = \frac{(C_{11} + C_{12} - 2C_{13})}{(C_{33} - C_{13})}$. A value of $f = 1$ indicates uniform atomic bonding in both the $a$- and $c$-orientations. An $f$-index below 1 signifies the *c*-direction is less compressible than the *a*-direction, while a value above 1 denotes enhanced compressibility along the *c*-direction compared to the *a*-direction. Among the compounds under investigation, $Mo_2Ta_2AlC_3$ is found to have the highest *f*-index value, indicating robust bonding in the horizontal plane (ab plane) (f > 1) and making it the most suitable candidate for the exfoliation process.

### 3.4.2 Machinability index and hardness

The performance of solids in the tribological sector is often estimated via the machinability index (MI) ($B/C_{44}$) [51]. Table 4 presents the MI index, which shows that the MI values for the GGA-PBE and GGA-PBEsol functions differ due to the varying values of $B$ and $C_{44}$. It is found that, in contrast to $Mo_2Zr_2AlC_3$ and $Mo_2Ta_2AlC_3$, $Mo_2Nb_2AlC_3$ exhibits greater machinability. Among the studied phases, $Mo_2Zr_2AlC_3$ depicts the smallest MI value, attributed to its elevated $C_{44}$, since materials with greater hardness generally offer reduced machinability. When addressing engineering applications, the hardness of a crystalline substance is a crucial consideration in manufacturing diverse systems. Hardness values can be derived from the elastic properties of polycrystalline materials since a material's resistance to indentation corresponds directly to its hardness. Both the micro-hardness ($H_{micro}$) and macro-hardness ($H_{macro}$) were computed employing Chen's and Miao's formula as follows $H_{macro} = 2[(\frac{G}{B})^2 G]^{0.585} - 3$ [52] and $H_{micro} = \frac{(1-2v)Y}{6(1+v)}$ [51]. Table 4 displays the hardness values of $Mo_2A_2AlC_3$ (A = Zr, Nb, Ta) alongside those of their corresponding o-MAX phase counterparts. The precision of the present assessment is validated by the resulting outcomes, which consistently align in correspondence with the stated values presented in Table 4. Materials with a hardness of greater than 10 GPa are outstanding for wear-resistant coatings and cutting tools, while those with a hardness of less than 10 GPa are better suited for applications that can be machined and withstand damage [53]. Therefore, it can be stated that the studied compounds are potential candidates for fabricating instruments and durable protective films combined with shielding layers under rigorous conditions. This is owing to their deformation tolerance, manufacturability, and moderate rigidity. A significant mechanical characteristic that evaluates a material's resistance to localized plastic deformation, like indentation or penetration, is a material's hardness. In crystalline solids, hardness is fundamentally governed by the atomic bonding and the underlying microstructural attributes.

MAX phases exhibit metal-like conductivity with ceramic-like stability owing to their unique nanolayered structure. Within these compounds, the M-A bonds tend to be weaker and more metallic in nature, while the M–X bonds are characterized by robust covalent and ionic interactions. Consequently, MAX phases exhibit moderate hardness, less than traditional ceramics but more than most metals, as well as improved machinability and damage tolerance. The experimentally determined hardness values are impacted by the specific measurement methods employed, whereas theoretical predictions of hardness also depend on the computational method used.

In this investigation, Vickers hardness has also been calculated as the geometric mean of all the bonds present through the crystalline structure. Gao et al. [54] initially introduced a methodology for estimating Vickers hardness based on Mulliken bond population analysis, particularly applicable to materials dominated by covalent bonding. Subsequently, Gou et al. [37] further improved this method to account for the partial metallic nature of bonding in compounds such as borides and carbides, making it more suitable for a broader class of materials and widely accepted in present theoretical studies [55]. The relevant formula for the Vickers hardness ($H_v$) based on the Mulliken bond populations is given as [56]:

$$H_v = \left[ \prod^{\mu} \left\{ 740 \left(P^{\mu} - P^{\mu'}\right) \left(v_b^{\mu}\right)^{\frac{-5}{3}} \right\}^{n^{\mu}} \right]^{\frac{1}{\sum n^{\mu}}} \qquad (6)$$

Here, $P^{\mu}$ is the Mulliken population of the $\mu$-type bond, $P^{\mu'} = n_{free} / V$ is the metallic population, and $v^{\mu}{}_b$ is the bond volume of the $\mu$-type bond, which is calculated in the following way:

$$v^{\mu}b = (d\mu)^3 / \sum v[(d\mu)^3 N^{\mu}b] \qquad (7)$$

Table 5 summarizes the computed Vickers hardness of $Mo_2A_2AlC_3$ (A = Zr, Nb, Ta), alongside that of $Mo_2Ti_2AlC_3$. The calculated values are 6.025 GPa, 6.277 GPa, and 6.224 GPa for $Mo_2Zr_2AlC_3$, $Mo_2Nb_2AlC_3$, and $Mo_2Ta_2AlC_3$, respectively. These values fall within the typical range reported for MAX phases, which usually show Vickers' hardness ($H_v$) values between 2 and 8 GPa [25]. $H_v$ values of related phases, such as $V_4AlC_3$ (6.74 ± 0.12 GPa) and $Mo_2Ta_2AlC_3$ (7.31 GPa), are also considered. Therefore, the $Mo_2A_2AlC_3$ (A = Zr, Nb, and Ta) 413 o-MAX phase compounds are ranked according to their $H_v$ values as follows: $Mo_2Nb_2AlC_3$ > $Mo_2Ta_2AlC_3$ > $Mo_2Zr_2AlC_3$.

**Table 5:** Calculated Mulliken bond number $n^\mu$, bond length $d^\mu$, bond overlap population $P^\mu$, metallic population $P^{\mu'}$, bond volume $v^\mu_b$, bond hardness $H^\mu_v$ of the $\mu$-type bond, and Vickers hardness $H_v$ of $Mo_2A_2AlC_3$ (A = Zr, Nb, Ta).

| Compounds | Bond | $n^v$ | $d^v$ (Å) | $P^\mu$ | $P^{\mu'}$ | $v^v_b(A^3)$ | $H^\mu_v$ (GPa) | $H_v$ (GPa) | Ref. |
|---|---|---|---|---|---|---|---|---|---|
| $Mo_2Zr_2AlC_3$ | C-Mo | 4 | 2.13271 | 1.32 | 0.02 | 14.753 | 10.10 | | This |
| | C-Zr1 | 4 | 2.28287 | 1.00 | 0.02 | 18.093 | 5.29 | 6.025 | |
| | C-Zr2 | 4 | 2.30893 | 0.65 | 0.02 | 18.721 | 3.03 | | |
| $Mo_2Nb_2AlC_3$ | C-Mo | 4 | 2.02945 | 1.01 | 0.02 | 14.406 | 8.20 | | This |
| | C-Nb1 | 4 | 2.15198 | 0.83 | 0.02 | 17.175 | 4.95 | 6.277 | |
| | C-Nb2 | 4 | 2.16569 | 0.91 | 0.02 | 17.505 | 5.30 | | |
| $Mo_2Ta_2AlC_3$ | C-Mo | 4 | 2.07463 | 0.52 | 0.03 | 13.950 | 3.90 | | This |
| | C-Ta1 | 4 | 2.27555 | 1.51 | 0.03 | 18.408 | 8.17 | 6.224 | |
| | C-Ta2 | 4 | 2.28173 | 1.14 | 0.03 | 18.559 | 5.95 | | |
| $Mo_2Ti_2AlC_3$ | C-Mo | 4 | 2.0980 | 1.21 | 0.03 | 14.67 | 9.92 | | [7] |
| | C-Ti1 | 4 | 2.1344 | 1.01 | 0.03 | 15.45 | 7.56 | 7.31 | |
| | C-Ti2 | 4 | 2.1447 | 0.72 | 0.03 | 15.67 | 5.20 | | |

### 3.4.3 The Brittleness of $Mo_2A_2AlC_3$ (A = Zr, Nb, Ta)

In this part of the study, three models were employed to determine the ductile or brittle behavior of $Mo_2A_2AlC_3$ (A = Zr, Nb, Ta). Ductility and brittleness are opposing phenomena, and their balance can be assessed through the shear-to-bulk modulus ratio, known as the Pugh ratio ($G/B$) [57]. For brittle solids, fracture is easier; on the other hand, plasticity is easier for ductile solids [58]. A $G/B$ ratio of 0.571 represents the threshold limit for evaluating the shift from ductile to brittle phases. Materials are ductile when their $G/B$ ratio does not surpass 0.571; otherwise, they are brittle. It is apparent in Table 2 that the $Mo_2A_2AlC_3$ (A = Zr, Nb, Ta) compounds are brittle substances. Poisson's ratio ($\upsilon$) serves as an alternative metric to define whether a substance behaves in a brittle or ductile manner. It can be obtained by applying the corresponding equation: $\upsilon = (3B - Y) / (6B)$ [59]. The brittle and ductile solids are separated by a Poisson's ratio of 0.26 by Frantsevich's criterion [60]. A value greater (or less) than 0.26 indicates ductility (or brittleness).

Similar to the *G/B* criterion, υ also confirms the brittle nature of the $Mo_2A_2AlC_3$ (A = Zr, Nb, Ta) compounds. Aside from brittleness, Poisson's ratio provides significant insights into the bonding interaction of crystalline materials. Covalent crystals usually exhibit a lower Poisson's ratio, around 0.10, while ionic crystals are associated with a higher value close to 0.33 [61]. As Table 4 demonstrates, the values of υ for $Mo_2A_2AlC_3$ (A = Zr, Na, Ta) exist between the two extremes, suggesting a mixture of ionic and covalent bonds. Additionally, when υ lies within the interval from 0.25 to 0.50, the dominant interatomic forces within the solids are central; otherwise, the interatomic forces are non-central [62]. $Mo_2A_2AlC_3$ (A = Zr, Nb, Ta) are dominated by non-central interatomic forces, given that the calculated values fall outside the specified range. Based on the value of Cauchy pressure, Pettifor [63] introduced an additional metric to analyze the bonding nature and fracture behavior of materials. The subsequent relationships may be adopted to calculate the *CP* from the elastic constants: $CP = C_{23}-C_{44}$, $C_{13}-C_{55}$, and $C_{12}-C_{66}$. A *CP* value greater than zero indicates materials with ductility, whereas a negative *CP* corresponds to materials exhibiting brittleness [58]. Yet again, the *CP* values for $Mo_2A_2AlC_3$ (A = Zr, Nb, Ta) characterize them as brittle materials. Furthermore, since the studied compounds have substantial negative values of *CP*, they should be dominated by directional covalent bonding.

### 3.4.4 Mechanical anisotropy

MAX phases have an abundance of scope for exploitation in cutting-edge technological implementations, supported by the extensive preceding research. Since mechanical anisotropy has a profound connection to crucial physical phenomena like micro-scale crack formation, defect mobility, permanent distortion in materials, phase segregation, and unusual phonon modes, a thorough understanding of it is therefore necessary across a variety of technological sectors [64]. Given its significance, we computed the two- and three-dimensional graphic models of the elastic moduli by using the ELATE program [65] along with the elastic anisotropy parameters of $Mo_2A_2AlC_3$ (A= Zr, Nb, Ta), which are provided in the Supplementary (Table 1 and Fig. 1-3, respectively).

This study also involved the computation of selected anisotropic indices. For example, the following relations have been employed for determining the directional shear anisotropy $A_i$ (i = 1-3) [66];

$$A_1 = \frac{\frac{1}{6}(C_{11} + C_{12} + 2C_{33} - 4C_{13})}{C_{44}}$$

$$A_2 = \frac{2C_{44}}{C_{11} - C_{12}}$$

$$A_3 = A_1 . A_2 = \frac{\frac{1}{3}(C_{11} + C_{12} + 2C_{33} - 4C_{13})}{C_{11} - C_{12}}$$

As demonstrated in Table 6, the calculated $A_i$ parameters deviate from unity, signifying the anisotropic mechanical response of the investigated phases. Utilizing the following relations, the anisotropy index for bulk modulus along $a$-and $c$-directions is calculated [57].

$$B_a = a\frac{dP}{da} = \frac{A}{\alpha} \; ; B_c = c\frac{dP}{dc} = \frac{B_a}{\alpha}, \text{ where } A = 2(C_{11} + C_{12}) + 4C_{13}\alpha + C_{33}\alpha^2 \text{ and } \alpha = \frac{(C_{11} + C_{12}) - 2C_{13}}{C_{33} + C_{13}}.$$

The values of $B_a$ and $B_c$ are non-equivalent, which indicates the anisotropy of the bulk modulus for $Mo_2A_2AlC_3$ (A = Zr, Nb, Ta). Furthermore, the linear compressibility coefficient ratio ($k_c/k_a$), is computed employing the following equations [67];

$$\frac{K_c}{K_a} = \frac{C_{11} + C_{12} - 2C_{13}}{C_{33} - C_{13}}$$ Since a value of $k_c/k_a$ =1, denotes the isotropic nature of solids thus, $k_c/k_a \neq 1$ for the studied compounds indicates the anisotropy of compressibility.

Table 6 presents the universal anisotropic index ($A^U$), computed based on the corresponding established relation [68,69]; $A^U = 5\frac{G_V}{G_R} + \frac{B_V}{B_R} - 6 \geq 0$. Since $A^U = 0$ for isotropic solids, an $A^U$ value of greater than zero suggests that $Mo_2A_2AlC_3$ (A = Zr, Nb, Ta) are anisotropic. Finally, based on Table 6, we can conclude that the $Mo_2A_2AlC_3$ (A = Zr, Nb, Ta) phases are anisotropic.

**Table 6:** Anisotropy factors of $MO_2A_2AlC_3$ (A = Zr, Nb, Ta).

| Phases | $A_1$ | $A_2$ | $A_3$ | $B_a$ | $B_c$ | $K_c/K_a$ | $A_B$ | $A_G$ | $A^U$ | Ref. |
|---|---|---|---|---|---|---|---|---|---|---|
| $Mo_2Zr_2AlC_3$ | 0.66 | 1.18 | 2.36 | 666.45 | 580.90 | 1.14 | 0.07 | 1.15 | 0.11 | This |
| $Mo_2Nb_2AlC_3$ | 0.85 | 0.97 | 2.51 | 760.55 | 581.07 | 1.30 | 0.32 | 0.29 | 0.03 | This |
| $Mo_2Ta_2AlC_3$ | 0.97 | 0.92 | 2.71 | 725.82 | 460.93 | 1.57 | 0.11 | 0.32 | 0.05 | This |
| $Mo_2V_2AlC_3$ | 0.99 | 0.89 | 0.89 | 791.66 | 732.31 | 1.08 | - | - | 0.02 | [14] |

### 3.5 Thermal properties

MAX phases exhibit exceptional mechanical performance under elevated thermal conditions, highlighting their suitability for deployment in thermally demanding environments. To gain a detailed comprehension of their thermodynamic response under equilibrium states, thermodynamic potential functions are evaluated by leveraging the PHDOS computed through the quasi-harmonic framework [69]. Several thermal traits can be used to understand a material's thermal behavior. Among the key solid-state parameters, the $\Theta_D$ facilitates understanding of certain fundamental physical aspects that bridge thermal and mechanical properties through crystal dynamics. For instance, the Debye temperature ($\Theta_D$) directly corresponds to vibrational enthalpy, melting temperature, heat capacities, thermal conductivity, phonons, and thermal expansions. A high ($\Theta_D$) value typically denotes a material with increased hardness. $\Theta_D$ is computed utilizing a well-established technique developed by Anderson et al. [70]:

$$\Theta_D = \frac{h}{k_B} \left[ \left( \frac{3n}{4\pi} \right) \frac{N_A \rho}{M} \right]^{\frac{1}{3}} v_m \qquad (8)$$

In this context, $M$ refers to the molar mass, $\rho$ denotes the corresponding density, $n$ signifies the number of atoms per molecule, while $h$, $k_B$, and $N_A$ stand for Planck's constant, Boltzmann's constant, and Avogadro's number, respectively. The following expression is employed to compute the mean sound velocity, ($v_m$):

$$v_m = \left[ \frac{1}{3} \left( \frac{1}{v_l^3} + \frac{2}{v_t^3} \right) \right]^{\frac{-1}{3}} \qquad (9)$$

Where $v_l$ and $v_t$ represent the longitudinal and transverse sound velocities, respectively, and are calculated employing the subsequent expressions:

$$v_l = \left[ \frac{(3B + 4G)}{3\rho} \right]^{\frac{1}{2}} \qquad (10)$$

$$v_t = \left[ \frac{G}{\rho} \right]^{\frac{1}{2}} \qquad (11)$$

Key physical quantities, including the density ($\rho$), Debye temperature ($\Theta_D$), mean atomic weight ($M/n$), along with various sound velocities ($v_l$, $v_t$, and $v_m$), are computed for the compounds under study, deploying GGA-PBEsol as illustrated in Table 7, along with $Mo_2Ti_2AlC_3$ and $Mo_2V_2AlC_3$ for comparison. According to our estimates, it is observed that the density has an inverse correlation with the average sound velocity ($v_m$). Traditionally, greater rigidity and elasticity are associated with an increased ($v_m$). A noteworthy trend observed in Table 7 is that the values of $v_l$ are significantly higher than those of $v_t$; compared to the longitudinal mode, the transverse mode experiences a reduction in wave velocity due to a greater loss of energy in vibrating nearby atoms [71]. As stated by Anderson et al. [70], the mean atomic weight (*M/n*) has a significant impact on the ($\Theta_D$). Table 7 suggests that when *M/n* increases, $\Theta_D$ decreases. Our findings are appropriately consistent with this concept. Amongst the compounds under investigation, $Mo_2Zr_2AlC_3$ has a higher Debye temperature ($\Theta_D = 634.09$ K) compared to $Mo_2Nb_2AlC_3$ ($\Theta_D = 632.60$ K) and $Mo_2Ta_2AlC_3$ ($\Theta_D = 525.98$ K), implying that $Mo_2Zr_2AlC_3$ is therefore superior for applications necessitating effective thermal management and state-of-the-art heat transfer elements, as it tends to have enhanced heat transfer capability and robust interatomic bonding. It is noteworthy that a MAX phase ($V_2SnC$) was recently reported as a TBC material with a $\Theta_D$ value of 472 K by Hadi et al. [72].

The thermal conductivity of a material, which is intrinsically related to the speed of sound in the material and PHDOS, indicates how effectively it conducts heat within the material. As the temperature rises, thermal conductivity diminishes and eventually reaches a lower bound in elevated-temperature regions [73]. Consequently, figuring out the minimum thermal conductivity plays a vital role in evaluating the viability of materials under extreme thermal conditions. The following modified Clarke's model [73] can be used to estimate a material's minimum thermal conductivity ($K_{min}$):

$$K_{min} = k_B . v_m \left( \frac{M}{n \rho N_A} \right)^{\frac{-2}{3}} \tag{9}$$

The meanings of the notations in this context retain the same definitions as those of the $\Theta_D$ expression. Table 7 summarizes the computed ($K_{min}$) values, revealing that $Mo_2Ta_2AlC_3$ has the lowest and $Mo_2Nb_2AlC_3$ has the highest among the evaluated compounds. Typically, thermal barrier coating (TBC) materials should exhibit a low $K_{min}$ in conjunction with other desirable thermal properties, namely a favorable *TEC* within a pronounced $T_m$, among others. With reduced ($K_{min}$) values, our findings demonstrate significant promise as TBC materials, in accordance with the attributes of substantiated TBC materials, which are described in the following section (3.6).

The Grüneisen parameter ($\gamma$), which measures the anharmonic effects in crystalline materials, has also been calculated using Poisson's ratio formula [63]:

$$\gamma = \frac{3}{2} \frac{(1 + \upsilon)}{(2 - 3\upsilon)} \tag{10}$$

Table 7 presents $\gamma$ values that fall within the expected range of 0.85 to 3.53 for polycrystalline materials with Poisson's ratios between 0.05 and 0.46 [74]. The low $\gamma$ values suggest that these compounds exhibit less anharmonic effects, which is consistent with stable lattice dynamics at various temperatures.

**Table 7**: Calculated Density($\rho$), sound velocities ($v_l$, $v_t$, and $v_m$), Debye temperature ($\Theta_D$), Gruneisen parameter ($\Upsilon$), minimum thermal conductivity ($K_{min}$), lattice thermal conductivity ($K_{ph}$), melting temperature ($T_m$), and thermal expansion coefficient (*TEC*) of $MO_2A_2AlC_3$ (A = Zr, Nb, Ta).

| Phases | $\rho$ (gm/cm$^3$) | $v_l$ (m/s) | $v_t$ (m/s) | $v_m$ (m/s) | $\Theta_D$ (K) | $\Upsilon$ | $K_{min}$ (W/mK) | $K_{ph}$ (W/mK) | $T_m$ (K) | *TEC* ($10^{-6}$ K$^{-1}$) | Ref. |
|---|---|---|---|---|---|---|---|---|---|---|---|
| $Mo_2Zr_2AlC_3$ | 7.04 | 7572 | 4520 | 3546 | 634 | 1.37 | 0.89 | 14.66 | 2064 | 11.11 | This |
| $Mo_2Nb_2AlC_3$ | 7.45 | 7551 | 4423 | 3472 | 632 | 1.44 | 0.90 | 12.98 | 2237 | 9.90 | This |
| $Mo_2Ta_2AlC_3$ | 10.05 | 6243 | 3731 | 2927 | 525 | 1.37 | 0.74 | 11.83 | 2131 | 11.47 | This |
| $Mo_2Ti_2AlC_3$ | 6.35 | 8361 | 4964 | 5497 | 726 | 1.40 | 1.49 | 12.30 | 2275 | - | [7] |
| $Mo_2V_2AlC_3$ | 6.76 | 8242 | 4737 | 5261 | 706 | 1.50 | 1.47 | 26.04 | 2432 | - | [14] |

The melting temperature ($T_m$) - a key parameter in predicting a material's performance at elevated temperatures was estimated for the studied compounds $Mo_2A_2AlC_3$ (A = Zr, Nb, and Ta) using the empirical method by Fine et al. [75][76] as follows: $T_m = 354 + 1.5(2C_{11} + C_{33})$, and is outlined in Table 7. The melting temperature of solids is predominantly governed by atomic bond strength, with higher $T_m$ values indicating stronger atomic bonding. Herein, the sequence of melting temperatures for the investigated phases corresponds closely with the order of Young's modulus, suggesting a strong interrelation between $T_m$ and $Y$. Our investigated compounds exhibit superior prospects for use as thermal barrier coatings, with high melting temperatures of 2064 K, 2237 K, and 2131 K for $Mo_2Zr_2AlC_3$, $Mo_2Nb_2AlC_3$, and $Mo_2Ta_2AlC_3$, respectively.

When a material is subjected to a temperature gradient, its lattice thermal conductivity ($K_{ph}$) measures the amount of heat transferred by atomic vibration. Materials with reduced $K_{ph}$ are often selected for insulating purposes, while those with intensified $K_{ph}$ are preferred in heat dissipation systems. For assessing the $K_{ph,}$ we applied the following empirical formula introduced by Slack [76]:

$$K_{ph} = A(\gamma) \frac{M_{av} \Theta_D^3 \delta}{\gamma^2 n^{\frac{2}{3}} T} \tag{11}$$

Here, $M_{av}$ denotes the average mass per atom, $n$ represents the total number of atoms within the unit cell, $T$ indicates the absolute temperature, $\gamma$ refers to the Grüneisen parameter, and A($\gamma$) is a corresponding function of $\gamma$. After computing the Grüneisen parameter, the coefficient A($\gamma$) can be evaluated from Juliann's equation as follows [77]:

$$A(\gamma) = \frac{4.85628 \times 10^7}{2 \left( 1 - \frac{0.514}{\gamma} + \frac{0.228}{\gamma^2} \right)} \tag{12}$$

A temperature-dependent analysis of $k_{ph}$ was conducted, extending up to 1100 K, and is presented in Fig. 7(d). A steady reduction in $k_{ph}$ is evident with rising temperatures and finally attains a saturation level at higher temperatures. Table 7 summarizes the lattice thermal conductivity ($K_{ph}$) of Mo$_2$A$_2$AlC$_3$ (A = Zr, Nb, Ta) o-MAX phases at 300 K (room temperature). At ambient temperature, all MAX phase compounds usually possess a lattice thermal conductivity range of 2.5 to 36 W/mK [78]. For the considered compounds, it is observed that Mo$_2$Zr$_2$AlC$_3$ exhibits the highest $K_{ph}$ of 14.66 W/mK, whereas Mo$_2$Ta$_2$AlC$_3$ exhibits a marginally lower value of 11.83 W/mK [see Table 7 and Fig. 7(d)]. These findings correspond to the anticipated range of lattice thermal conductivity for MAX phases. Moreover, the observed trend of decreasing $K_{ph}$ with temperature indicates strong phonon scattering, which is highly desirable for TBC applications where thermal insulation must be maintained under extreme thermal loads. The reported conductivity values suggest that these phases offer an effective compromise between low heat transfer and mechanical integrity, reinforcing their potential utility as durable high-temperature barrier materials

To evaluate the heat capacity ($C_v$) over the 0-1000 K range, the quasi-harmonic Debye scheme was employed as follows [79] [58]:

$$C_v = 9nN_A k_B \left( \frac{T}{\Theta_D} \right) \int_0^{x_D} dx \ \frac{x^4}{(e^x - 1)^2} \tag{13}$$

Wherein, $x_D = \frac{\Theta_D}{T}$. The following equations were employed to estimate the linear thermal expansion coefficient (TEC) and the specific heat at constant pressure ($C_P$) [79]:

$$\alpha = \frac{\gamma C_v}{3 B_T v_m} \tag{14}$$

$$C_p = C_v(1 + \alpha\gamma T) \qquad\qquad\qquad (15)$$

Where $B_T$ denotes isothermal bulk modulus, $v_m$ represents molar volume, and $\gamma$ corresponds to the Grüneisen parameter. Figure 7(a, c) displays the temperature dependence of $C_v$, $C_p$, and α for the investigated compounds, which were calculated using the equations above. As the temperature increases, phonon softening becomes more pronounced, resulting in a higher heat capacity. The heat capacities are noticed to increase significantly with temperature up to 300 K, following the Debye-$T^3$ model [80], and at the elevated temperature regime nearly approach the Dulong-Petit ($3nNAkB$) limit, where the $C_v$ and $C_p$ do not intensely rely on the temperature [81].

Anharmonic effects within the crystal lattice are responsible for thermal expansion, as reflected in the *TEC*, and this behavior is closely associated with the discrepancy between $C_v$ and $C_p$. Figure 7(c) depicts the evaluated values of *TEC* for $Mo_2A_2AlC_3$ (A = Zr, Nb, Ta) phases. A rapid escalation in *TEC* is observed up to 250 K; thereafter, it increases moderately between 250 K and 600 K and ultimately approaches a saturation state. The computed values of *TEC* are found to be $11.11\times10^{-6}$ K$^{-1}$, $9.90\times10^{-6}$ K$^{-1}$, and $11.47\times10^{-6}$ K$^{-1}$ for $Mo_2Zr_2AlC_3$, $Mo_2Nb_2AlC_3$, and $Mo_2Ta_2AlC_3$ compounds at a temperature of 300 K, respectively. The relatively high and stable *TEC* values at elevated temperatures indicate that these compounds can accommodate thermal stress more effectively, which is a critical requirement for reliable thermal barrier coating (TBC) applications.

Fig. 7(b) illustrates the thermodynamic functions, including Helmholtz free energy *F*, internal energy *E*, and entropy *S* for $Mo_2A_2AlC_3$ (A = Zr, Nb, Ta) compounds, derived at zero pressure using the quasi-harmonic approximation. As a natural process advances, free energy gradually decreases, intensifying a further downward trend. As the temperature rises, internal energy (*E*) increases, contrary to the decreasing trend observed in free energy. Thermal agitation intensifies the disorder within the system as the temperature increases, thereby causing an increase in entropy (S), as depicted in Fig. 7(b).

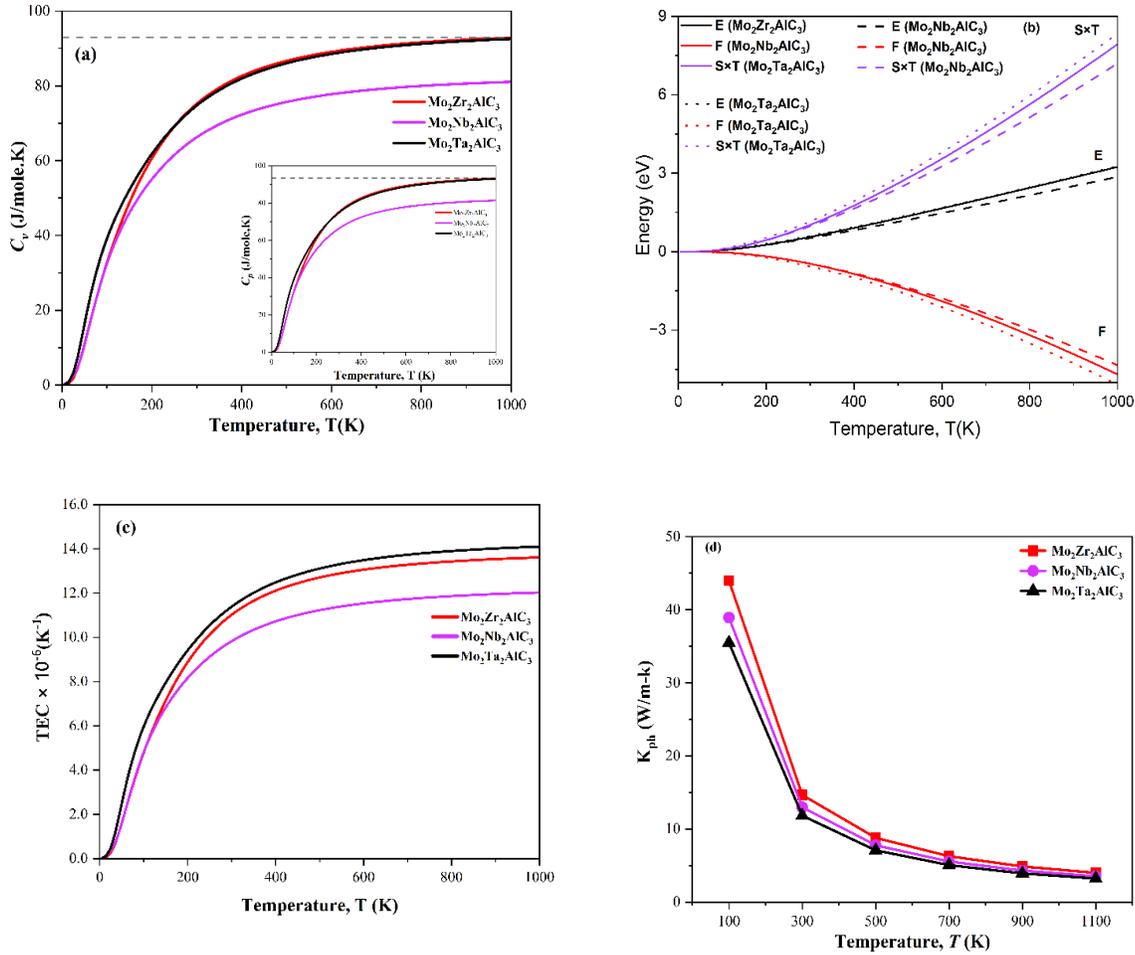

**Fig. 7**: Temperature dependence of the calculated thermodynamic parameters. (a) Specific heat $C_v$ and $C_p$ (b) Thermodynamic potential functions (*F, E, T.S),* (c) Thermal expansion coefficient (*TEC*), of Mo$_2$A$_2$AlC$_3$ (A = Zr, Nb, Ta).

### 3.5.1 Assessment of Thermal Barrier Coatings Performance Perspectives:

Modern engine design requires coating materials that can withstand intense functional domains while facilitating superior efficiency and reduced emissions. MAX phases offer a distinctive advantage as thermal barrier coatings (TBC), attributable to their layered structure and novel composition of tolerance to mechanical stresses and thermal resilience.

Accomplishing steady thermal endurance in TBCs requires materials that integrate substantial mechanical resilience with elevated hardness and fracture resistance, ensuring they can counteract thermal stress without degradation. Fracture toughness (*K$_{IC}$*) provides a quantitative assessment of how effectively a material impedes the propagation of cracks. Amplified fracture toughness in TBCs is pivotal to oppose stress-induced cracking, enabling the material to endure intense thermal cycling while preserving structural integrity and operational performance. In the context of proven TBCs, yttria-stabilized zirconia (YSZ) possesses a $K_{IC}$ value of 2.0-3.3 MPa.m$^{1/2}$, followed by lanthanum zirconate (LZ) at 1.1 ± 0.2 MPam$^{1/2}$ [82], while the prospective candidate SrZrO$_3$ demonstrates an intermediate value of 1.5 ± 0.1 MPam$^{1/2}$ [83]. Building upon the $K_{IC}$ trends observed in proven and prospective TBCs, our investigated compounds herein reveal prominent endurance to crack extension, with $K_{IC}$ values of 2.67 Mpam$^{1/2}$, 2.78 Mpam$^{1/2}$, and 2.59 Mpam$^{1/2}$ for Mo$_2$A$_2$AlC$_3$ phases (A = Zr, Nb, Ta) respectively, positioning these materials as strong contenders for thermal-intensive environments. Hardness evaluation of coating materials is typically performed through indentation testing across the nano, micro, and macro scales. Micro-indentation measurements revealed hardness values of 9.9 ± 0.4 GPa and 13 ± 1 GPa for lanthanum zirconate (LZ) and yttria-stabilized zirconia (YSZ), respectively [83]. For the investigated Mo$_2$A$_2$AlC$_3$ phases (A = Zr, Nb, Ta), the hardnesses calculated using Chen's formula [51] are 26.57 GPa, 25.40 GPa, and 25.94 GPa, respectively. The corresponding Vickers hardnesses are 6.02 GPa, 6.27 GPa, and 6.22 GPa, respectively. Although no direct comparison between theoretical and experimental values is possible, the hardness of the herein predicted compounds is expected to be sufficient for selecting them as TBC materials.

The minimum lattice thermal conductivity ($K_{min}$) plays a crucial role in TBC design, as materials with lower $K_{min}$ values reduce thermal diffusion, thereby enhancing the coating's protective effectiveness under extreme operating temperatures. A material can be interpreted as a prospective TBC candidate if its $K_{min}$ value falls below 1.25 W/m·K [84]. Among established TBCs, 3YSZ, CeO$_2$, mullite, Al$_2$O$_3$ and TiO$_2$ demonstrate $K_{min}$ values of 2.12 W/m·K (at 1273 K), 2.77 W/m·K (at 1273 K), 5.3 W/m·K (at 1273 K), 5.8 W/m·K (at 1400 K) and 3.3 W/m·K (at 1400 K) [85] [86]. Pyrochlore oxides of the form RE$_2$T$_2$O$_7$, incorporating rare-earth elements such as La, Pr, Nd, Sm, Eu, Gd, Y, Er, Lu at the RE site and Ti, Mo, Sn, Zr, or Pb at the T site, are considered reliable TBC candidates that exhibit $K_{min}$ values estimated between 1.40 and 3.05 W/m·K [80]. Perovskite oxides with the general formula ABO$_3$, where A-sites are occupied by Sr, Ba, and the B-sites by Ti, Zr, Hf, have been extensively utilized in TBC applications, depicting $K_{min}$ values that range from 1.09 to 1.74 W/m·K [80]. Within the family of rare-earth stannates formulated as RE$_2$Sn$_2$O$_7$, where La, Nd, Sm, Gd, Er, and Yb occupy the RE site, the reported $K_{min}$ values extend from 1.80 to 2.50 W/m·K [80]. There are several experimentally validated TBC systems that manifest low $K_{min}$ values, such as 1.22 W/m·K for $Gd_2Zr_2O_7$, 1.3 W/m·K for $Y_2SiO_5$, 1.20 W/m·K for $V_2SnC$, and 1.13 W/m·K for $Y_4Al_2O_9$ [87] [72] [88]. Herein investigated $Mo_2A_2AlC_3$ phases (A = Zr, Nb, Ta) display exceptionally low $K_{min}$ values of 0.89 W/m·K, 0.90 W/m·K, and 0.74 W/m·K, respectively, underscoring their potential as advanced TBC capable of competently suppressing heat propagation in extreme operating environments.

When designing thermal barrier coatings, the material's thermal expansion coefficient (*TEC*) plays a vital role in ensuring that the material's dimensional changes under high temperatures align with the substrate, thereby limiting crack formation and spallation. To illustrate typical *TEC* benchmarks, widely used TBC materials such as 3YSZ, $La_2Zr_2O_7$, $BaZrO_3$, 8YSZ, $Al_2O_3$ (TGO), NiCoCrAlY and IN737 superalloy display *TEC* values of $11.5 \times 10^{-6}$ $K^{-1}$, $9.1 \times 10^{-6}$ $K^{-1}$, $8.1 \times 10^{-6}$ $K^{-1}$, $10.7 \times 10^{-6}$ $K^{-1}$, $8.6 \times 10^{-6}$ $K^{-1}$, $17.5 \times 10^{-6}$ $K^{-1}$ and $16 \times 10^{-6}$ $K^{-1}$, sequentially [83] [89]. These benchmark values offer an effective baseline for assessing advanced coating systems. Building on these reference benchmarks, the investigated $Mo_2A_2AlC_3$ phases (where A corresponds to Zr, Nb, Ta) demonstrate *TEC* values of $11.11 \times 10^{-6}$ $K^{-1}$, $9.90 \times 10^{-6}$ $K^{-1}$, and $11.47 \times 10^{-6}$ $K^{-1}$ in sequence, highlighting their alignment with proven TBC material requirements.

Another critical aspect of a promising TBC material is its high melting temperature ($T_m$), which ensures thermal stability under rigorous conditions. To contextualize thermal endurance, accredited TBC materials including $La_2Zr_2O_7$, $BaZrO_3$, 3YSZ, $TiO_2$, Garnet ($Y_3Al_5O_{12}$), $LaPO_4$, $V_2SnC$, and $Y_4Al_2O_9$ featuring $T_m$ of 2573 K, 2963 K, 2973 K, 2098 K, 2243 K, 2343 K, 1533 K, and 2000 K, correspondingly, reinforcing their competence for scenarios where sustained high-temperature efficiency is decisive [90] [91] [92] [88]. Extending this comparison to the examined $Mo_2A_2AlC_3$ phases (A = Zr, Nb, Ta), the quantified $T_m$ values of 2064 K, 2237K, and 2131 K, in sequence, underscore their robust thermal resilience for TBC applications, where prolonged exposure to extreme heat is crucial.

The Grüneisen parameter ($\gamma$), functioning as a descriptor of lattice anharmonicity, plays a central role in governing thermal expansion mismatch and phonon–phonon scattering. For $Mo_2A_2AlC_3$ (A = Zr, Nb, Ta) phases, the values of $\gamma$ (1.37, 1.44, 1.37 in sequence) are close to that of proven TBC material $Y_4Al_2O_9$ (1.55), signifying regulated anharmonicity [88]. Such behavior implies that thermal strain within the coating can be effectively managed, reducing crack propagation under cyclic heating.

The overall assessment highlights $Mo_2A_2AlC_3$ (A = Zr, Nb, Ta) phases as promising candidates for designing next-generation TBCs capable of attaining extreme thermal demands.

## 3.5 Optical Properties:

The correlation of crystalline solids with electromagnetic radiation is crucial not only for fabricating optoelectronic devices but also for probing their optical features. They offer insights into the functional behavior of advanced materials and their potential applications in cutting-edge engineering infrastructures. To quantify the optical interactions of these novel o-MAX phases $Mo_2A_2AlC_3$ (A = Zr, Nb, Ta), we assessed various optical features, including the dielectric constant, refractive index, absorption coefficient, photoconductivity, reflectivity, and energy loss function across photon energies up to 30 eV. But only the dielectric function and reflectivity are presented here; the rest of the optical properties are provided in the supplementary section [Fig. 4 (a,b), Fig. 5(b), Fig. 6(a,b)] The material's optical behavior is described by the complex dielectric function, $\epsilon(\omega) = \epsilon_1(\omega) + i\epsilon_2(\omega)$, where $\omega$ is the angular frequency of the incident electromagnetic radiation [25]. The imaginary part, $\epsilon_2(\omega)$, quantifies light absorption due to electronic transitions and is calculated using:

$$\epsilon_2(\omega) = \frac{2e^2\pi}{\Omega\epsilon_0} \sum_{k,v,c} |\psi_k^c|u.r|\psi_k^v|^2 \delta(E_k^c - E_k^v - E) \qquad (16)$$

Here, $e$ is the electron charge, $\Omega$ is the volume of the unit cell, $\epsilon_0$ represents the permittivity of free space, the symbols $\psi_k^c$ and $\psi_k^v$ refer to the conduction band and valence band wavefunctions at a given wave vector $k$, $u$ indicates the polarization vector, and $\delta$ ensures energy conservation. The real component, $\epsilon_1(\omega)$, which describes the material's polarization response, can be obtained via the Kramers–Kronig transformation [25]:

$$\epsilon_1(\omega) = 1 + \frac{2}{\pi} P \int_0^\infty \frac{\omega' \epsilon_2(\omega')}{\omega'^2 - \omega^2} \, d\omega' \qquad (17)$$

Here, the principal value of the integral is indicated by $P$. Based on the dielectric function, secondary optical parameters, reflectivity $R(\omega)$, absorption coefficient $\alpha(\omega)$, refractive index $n(\omega)$, and extinction coefficient $k(\omega)$, are determined through well-established relations [93]:

$$R(\omega) = \left| \frac{\sqrt{\epsilon(\omega)} - 1}{\sqrt{\epsilon(\omega)} + 1} \right|^2 \qquad (18)$$

$$\alpha(\omega) = \frac{2\omega k(\omega)}{c} \qquad\qquad (19)$$

We calculated the optical properties of $Mo_2A_2AlC_3$ (A = Zr, Nb, Ta) through density functional theory (DFT) simulations carried out with CASTEP code. Both GGA-PBE and GGA-PBEsol exchange correlation functionals were utilized; however, the GGA-PBEsol results were chosen for the final analysis due to their improved reliability in describing the electronic and optical characteristics of MAX phases [16]. A Drude correction was included to account for the metallic character, employing an unscreened plasma frequency of 3.0 eV and a damping constant of 0.05 eV. In addition, to improve k-point convergence near the Fermi level, a Gaussian smearing of 0.5 eV was also used [53]. We investigated the optical properties considering polarization along the [100] and [001] crystallographic directions to examine the material's anisotropic optical response.

The dielectric function is crucial for explaining the material's response to incident photons [25]. Real part of the dielectric function, $\varepsilon_1(\omega)$, of $Mo_2A_2AlC_3$ (A = Zr, Nb, Ta) for [100] and [001] polarization directions is shown in Fig. 8(a). The real part, $\varepsilon_1(\omega)$, crosses zero from negative values for both polarization directions, at the low photon energy region ($\sim 0.5$eV), indicating consistency with the established Drude model for metallic materials. Within the $0.5 - 2.0$ eV energy range, $Mo_2Nb_2AlC_3$ and $Mo_2Ta_2AlC_3$ exhibit significant anisotropy. Particularly, $Mo_2Ta_2AlC_3$ shows pronounced anisotropy, where the [001] polarization direction exhibits lower zero crossing energy ($\sim 0.46$eV) and higher peak intensity compared to the [100] direction.

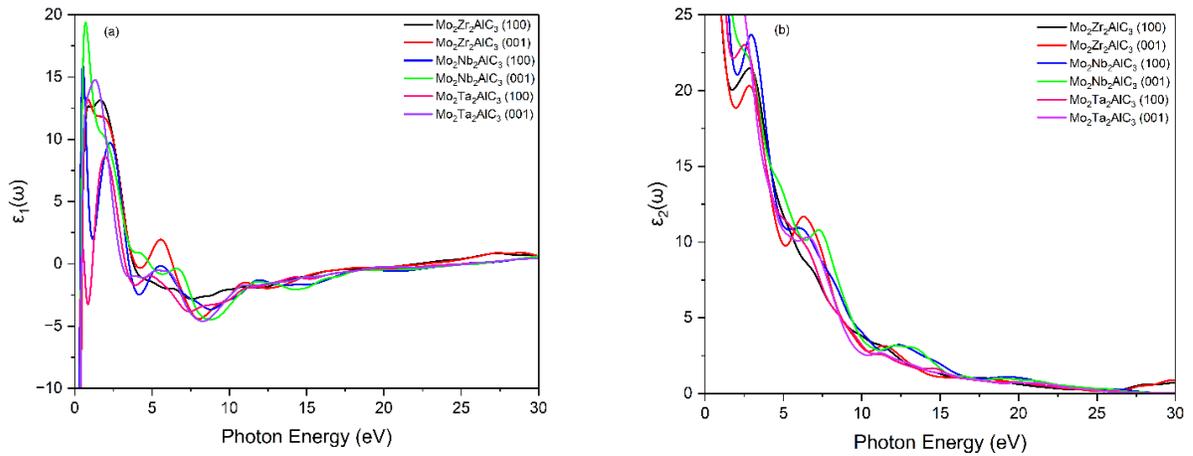

Fig. 8: The (a) real and (b) imaginary parts of the dielectric function.

The imaginary part of the dielectric function, $\varepsilon_2(\omega)$, is a crucial parameter for analyzing the optical characteristics of materials [25]. The calculated $\varepsilon_2(\omega)$ for both polarization directions is depicted in Fig. 8(b). The imaginary part $\varepsilon_2(\omega)$ shows a monotonic decrease with increasing photon energy for all three compounds, beginning with high values at low energies (~0.01eV) and gradually approaching zero at higher energies (~28.5eV), which confirms the metallic character of $Mo_2A_2AlC_3$ (A = Zr, Nb, Ta). The prominent peaks appearing at low photon energy range (~1eV) are mainly due to intra-band transitions, where free electrons from the same energy band are responsible for optical absorption. The [100] polarization direction for all compounds demonstrates higher $\varepsilon_2(\omega)$ values than the [001] direction across the majority of the energy range, which indicates stronger optical absorption in the [100] direction.

Reflectivity, $R(\omega)$, is the ratio of reflected energy to incident energy at angular frequency $\omega$, describing the surface's ability to reflect electromagnetic radiation [58]. The computed reflectivity $R(\omega)$ for $Mo_2A_2AlC_3$ (A = Zr, Nb, Ta) is presented in Fig. 9(a). In the low energy infrared region (~1eV), all compounds for both polarizations exhibit a high value of 0.98 (98%), indicating strong reflective behavior at this energy level. The reflectivity decreases significantly in the visible light region (1.8−3.1 eV). Reflectivity of $Mo_2Zr_2AlC_3$ and $Mo_2Nb_2AlC_3$, respectively, for the [100] and [001] polarization drops to around 50−53%, while $Mo_2Ta_2AlC_3$ shows a slightly higher value, around 54−55% in both planes. All compounds show stable reflectivity in the *UV* region (3.1−10 eV), $Mo_2Zr_2AlC_3$ maintains values around 42−49% for both planes, while $Mo_2Nb_2AlC_3$ and $Mo_2Ta_2AlC_3$ exhibit higher reflectivity, ranging from 49−51% in the [100] plane and 51−53% in the [001] plane. Studies have shown that MAX phases with reflectivities exceeding 44% across the visible to near-*UV* wavelength range are promising for use as coating materials [94]. Our results reveal that all examined compounds demonstrate the required reflective properties, indicating their potential as effective coating materials for thermal barrier applications.

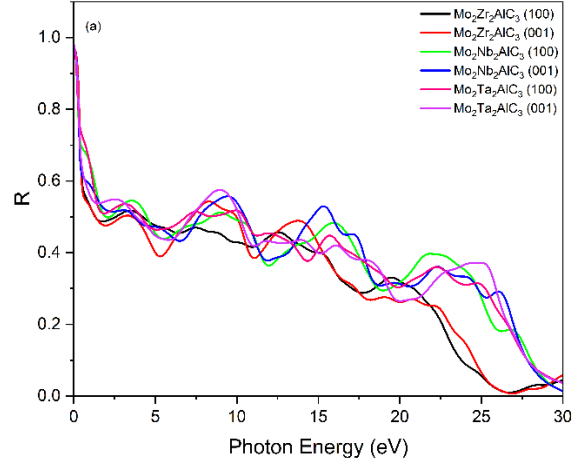

**Fig. 9:** (a) Reflectivity R($\omega$).

## 4. Conclusions

In this study, using the DFT approach, we explored three novel o-MAX phases, $Mo_2A_2AlC_3$ (A = Zr, Nb, Ta), and investigated their structural, lattice dynamical, electronic, mechanical, thermodynamic, and optical properties to predict their potential applications. Phonon spectra, formation energy, and elastic parameters collectively confirm the dynamic, chemical, mechanical, and thermodynamic stability of the studied phases. Both the electronic band structure and DOS certify the metallic nature, with prominent contributions from the Mo-4d states. The analysis of partial DOS together with CDM reveals the presence of mixing of ionic, covalent, and metallic bonding. Based on the evaluated elastic parameters, it can be inferred that these materials can endure applied forces without any structural distortion, which reflects their superior mechanical stability. Analysis of $\upsilon$, $G/B$, as well as CP confirms the brittleness of $Mo_2A_2AlC_3$ (A = Zr, Nb, Ta). The anisotropic nature of the examined phases is confirmed by directional variations in stiffness moduli and anisotropy indices. These compounds exhibit Vickers hardness values within the expected ranges, indicating their softness and ease of machinability. It is noted that $Mo_2Zr_2AlC_3$ exhibits higher damage tolerance due to its low $M_B$ value, whereas $Mo_2Nb_2AlC_3$ shows enhanced resistance to crack propagation. Key thermal characteristics, including *TEC, $T_m$, $\Upsilon$, $\Theta_D$, $k_{min}$, $k_{ph}$*, along with $C_v$, $C_p$, were determined. The small values of and *k*min, combined with favorable values of *TEC*, and the high values $T_m$ and $\Theta_D$, imply their promising potential as TBC materials in high-temperature applications. The optical constants' spectra suggest the intrinsic metallic behavior of the studied phases. The studied phases are potential candidates for solar heat-shielding layers, as evident from reflectivity values surpassing 44% across the IR, visible, and mid-UV spectra. To sum up, the outlined findings are expected to offer valuable insights into subsequent practical and theoretical evaluations regarding this captivating novel, the o-MAX phase.

## Acknowledgment

The authors are grateful to the Ministry of Science and Technology (MoST), Bangladesh, for providing financial support under Special Allocation for Science and Technology [Project ID: SRG:246460]. The authors are also grateful to the ACMRL lab of the Department of Physics, Chittagong University of Engineering & Technology (CUET), Chattogram-4349, Bangladesh, for providing the computing facility for this work.

## CRediT Author contributions

**M. I. A. Tanim:** Data curation, Investigation, Visualization, Formal analysis, Writing – original draft. **C. Talukder**: Data curation, Visualization, Writing – original draft. **S. S. Saif:** Data curation, Visualization, Writing – original draft. **Labib H. K. Adnan**: Data curation, Investigation, Visualization, Formal analysis. **N. Jahan:** Formal analysis, Validation, Writing - review & editing. **M. M. Hossain:** Formal analysis, Validation, Writing - review & editing. **M. M. Uddin**: Validation, Writing- review & editing. **M. A. Ali:** Conceptualization, Methodology, Formal analysis, Validation, Project administration, Writing – review & editing, Supervision.

## Data Availability statement:

The data will be made available on reasonable request.

**DFT prediction of new o-MAX phases: Mo$_2$A$_2$AlC$_3$ (A = Zr, Nb, Ta) for next-generation thermal barrier coatings**

M. I. A. Tanim, C. Talukder, S. S. Saif, Labib H. K. Adnan, N. Jahan, M. M. Hossain, M. M. Uddin, M. A. Ali*

Advanced Computational Materials Research Laboratory (ACMRL), Department of Physics

Chittagong University of Engineering and Technology (CUET), Chattogram-4349, Bangladesh

**Table 1:** Calculated minimum and maximum values of the Young's Modulus ($Y$ in GPa), linear compressibility ($K$ in TPa$^{-1}$), shear modulus ($G$ in GPa), and Poisson's ratio of Mo$_2$A$_2$AlC$_3$ (A = Zr, Nb, Ta).

| Phases | $Y_{min}$ | $Y_{max}$ | $A_Y$ | $K_{min}$ | $K_{max}$ | $A_K$ | $G_{min}$ | $G_{max}$ | $A_G$ | $\upsilon_{min}$ | $\upsilon_{max}$ | $A_\upsilon$ | Ref. |
|---|---|---|---|---|---|---|---|---|---|---|---|---|---|
| Mo$_2$Zr$_2$AlC$_3$ | 275 | 379 | 1.37 | 1.500 | 1.72 | 1.14 | 117 | 167 | 1.42 | 0.11 | 0.31 | 2.71 | This |
| Mo$_2$Nb$_2$AlC$_3$ | 304 | 374 | 1.23 | 1.314 | 1.72 | 1.30 | 132 | 152 | 1.15 | 0.20 | 0.29 | 1.42 | This |
| Mo$_2$Ta$_2$AlC$_3$ | 289 | 369 | 1.27 | 1.377 | 2.16 | 1.57 | 134 | 147 | 1.09 | 0.18 | 0.25 | 1.35 | This |
| Mo$_2$V$_2$AlC$_3$ | 359 | 145 | 1.11 | 1.26 | 1.37 | 1.08 | 145 | 163 | 1.13 | 0.22 | 0.28 | 1.31 | [1] |

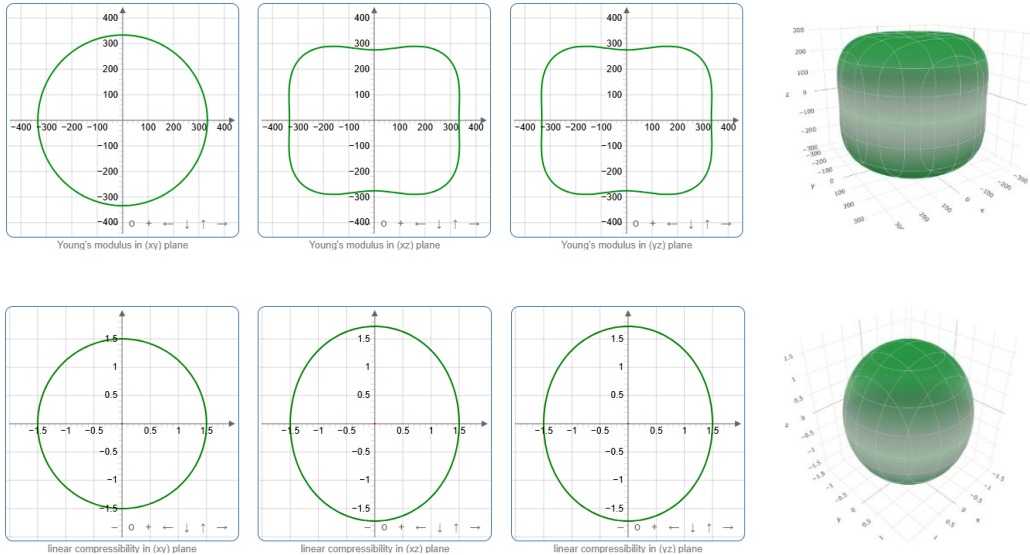

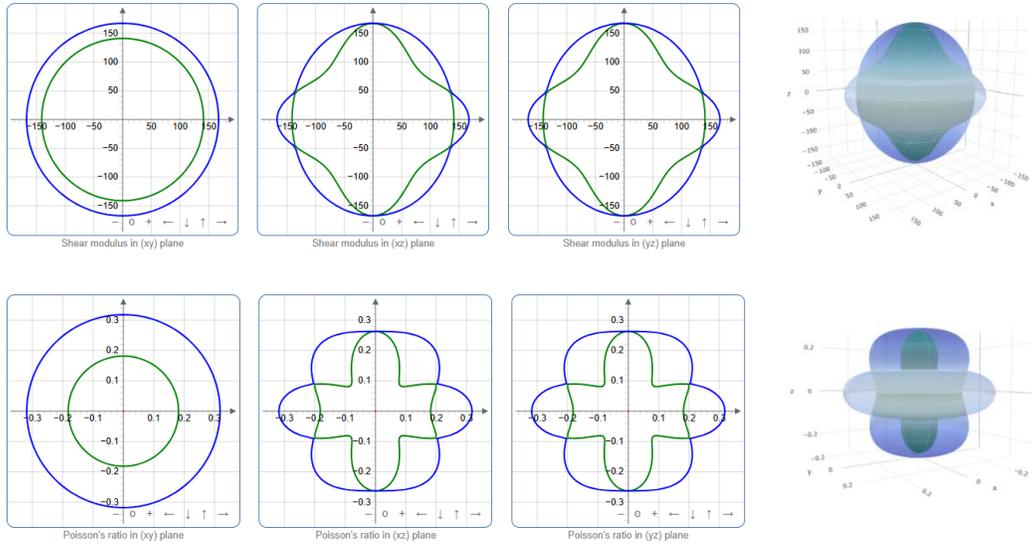

**Fig. 1**: The 2D and 3D plots of (a) *Y*, (b) *K*, (c) *G*, and (d) *v* of Mo$_2$Zr$_2$AlC$_3$.

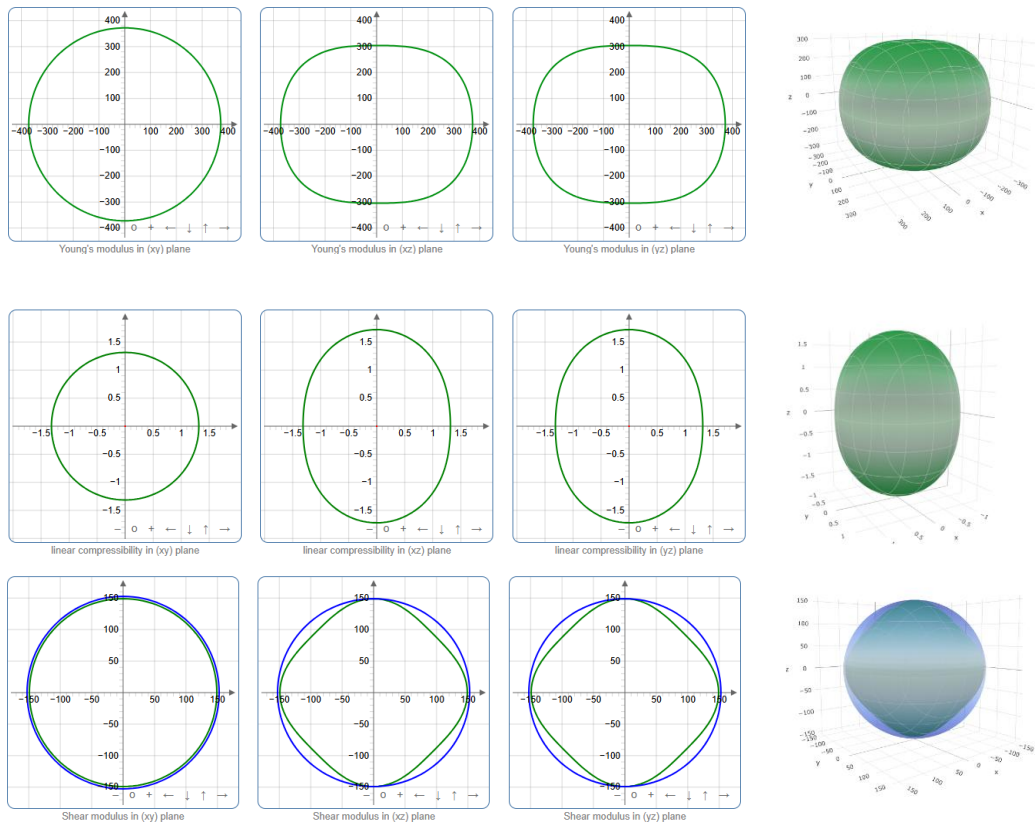

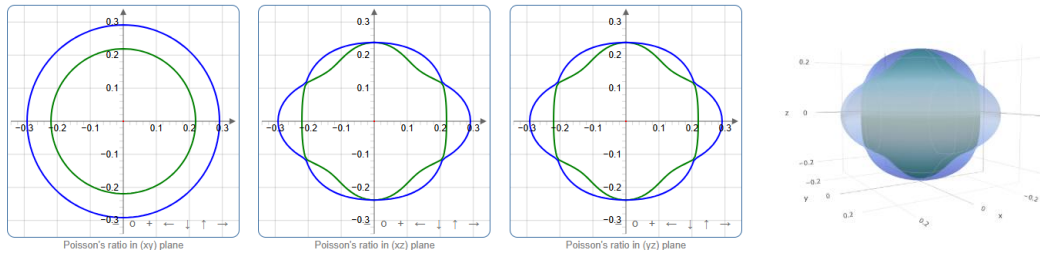

**Fig. 2**: The 2D and 3D plots of (a) *Y*, (b) *K*, (c) *G*, and (d) *v* of Mo$_2$Nb$_2$AlC$_3$.

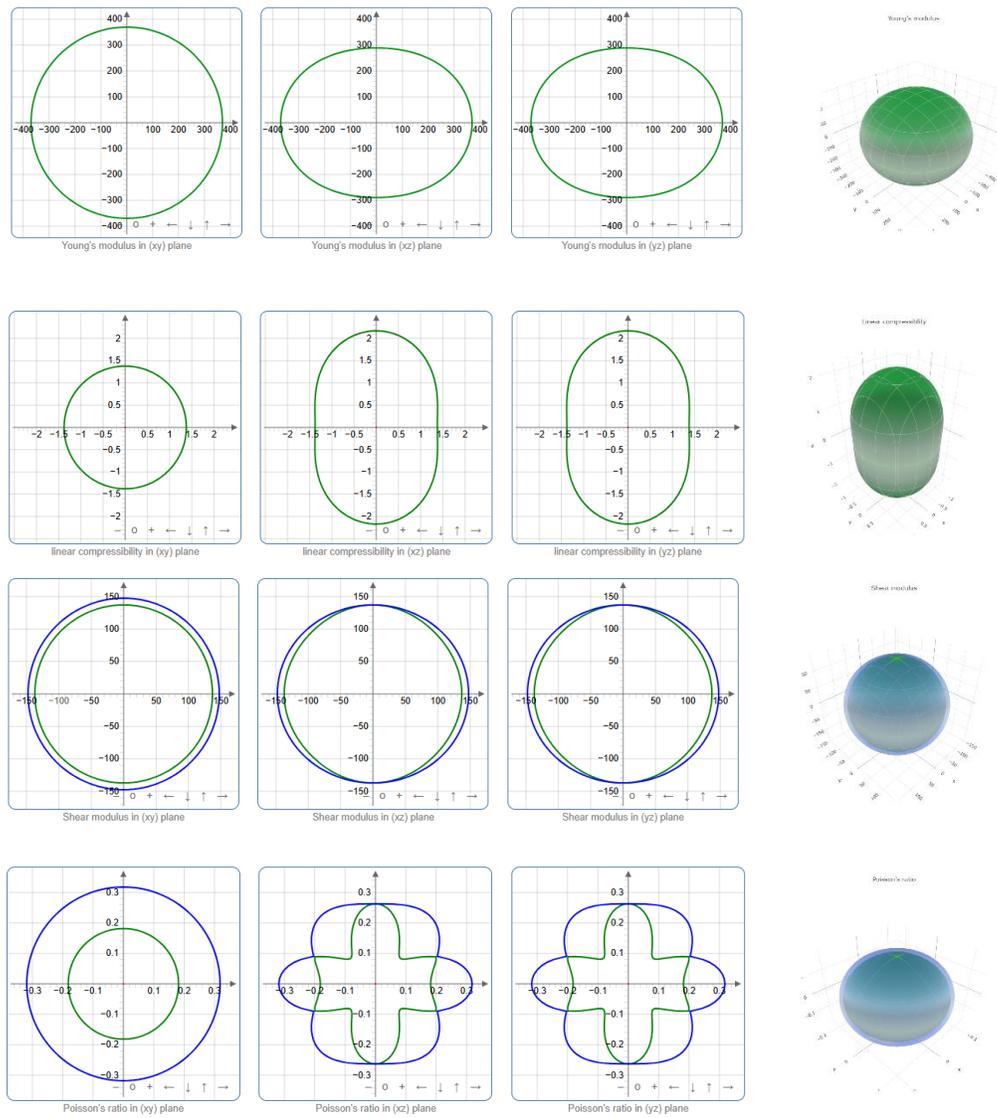

**Fig. 3**: The 2D and 3D plots of (a) *Y*, (b) *K*, (c) *G*, and (d) *v* of Mo$_2$Ta$_2$AlC$_3$.

**Optical properties**

Understanding the refractive index $n(\omega)$ of materials is crucial, as it informs about a material's optical behavior and interaction with electromagnetic radiation [2]. This information is essential for evaluating the suitability of these materials for diverse optical applications [2]. The refractive index $n(\omega)$ of $Mo_2A_2AlC_3$ (A = Zr, Nb, Ta), respectively, for the [100] and [001] polarization is shown in Fig. 4(a). At low photon energy (~0.01eV), the static values of $n(0)$ are high for all compounds, with values ranging from 84.47−85.12 across both planes, indicative of strong electronic polarization due to intra-band transitions of electrons. For $Mo_2Zr_2AlC_3$, $Mo_2Nb_2AlC_3$, and $Mo_2Ta_2AlC_3$, respectively, for the [100] ([001]) polarization, the static values of $n(0)$ are 84.50 (84.47), 84.84 (84.58), and 85.12 (84.66). The spectra of $n(\omega)$ for all three compounds in both planes decrease sharply within the photon energy range of 0.05–10 eV and attain their lowest values around 0.74–1.06. The refractive index $n(\omega)$ stabilizes at 10 eV with minor increases at higher energies.

The absorption loss of an electromagnetic wave while traversing a material can be measured by the extinction coefficient $k(\omega)$ [3] . At low photon energy (~0.01eV), all three compounds exhibit high extinction coefficients, which is shown in Fig. 4(b). For $Mo_2Zr_2AlC_3$, $Mo_2Nb_2AlC_3$, and $Mo_2Ta_2AlC_3$, respectively, for the [100] ([001]) polarization, show values of 102.52 (102.54), 102.30 (102.46), and 102.17 (102.43). As photon energy increases, $k(\omega)$ for all compounds decreases sharply. In the 2–10 eV range (visible to *UV*), the extinction coefficient stabilizes around 1.83–2.24 for all compounds, with local maxima corresponding to zero crossings of $\varepsilon_1(\omega)$.

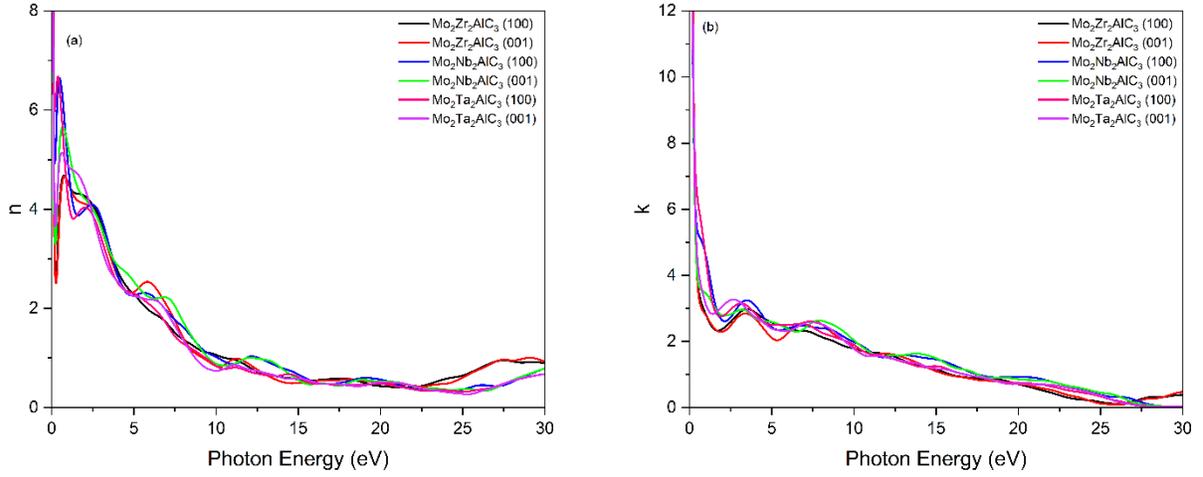

**Fig. 4**: (a) Refractive index n(ω) and (b) extinction coefficient k(ω).

The energy loss function $L(\omega)$, shown in Fig. 5(b), characterizes the energy dissipation of high-velocity electrons passing through the material [4]. It corresponds with bulk plasma frequency ($\omega_p$), at which the real part of the dielectric function approaches zero and the imaginary part stays below one. For both [100] and [001] polarization directions, the plasma frequency ($\omega_p$) is 20.98 eV and 21.49 eV for $Mo_2Zr_2AlC_3$; 20.68 eV and 20.98 eV for $Mo_2Nb_2AlC_3$, 20.98 eV and 21.19 eV for $Mo_2Ta_2AlC_3$, indicating slightly higher $\omega_p$ in the [001] planes for all the compounds. As the photon energy exceeds the plasma frequency $\omega_p$ (~21 eV), the energy loss function increases sharply, which indicates a transition where the absorption coefficient and reflectivity decrease rapidly. At frequencies higher than this critical value, the material becomes transparent, allowing electromagnetic fields to propagate through it with minimal loss.

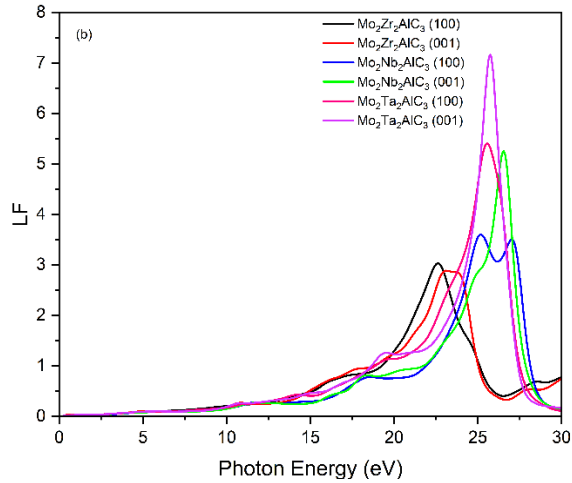

Fig. 5: (b): energy loss function L($\omega$)

The absorption coefficient $\alpha(\omega)$ quantifies how effectively a material absorbs incident electromagnetic radiation across the spectrum, providing key information on its suitability for solar-energy applications [5]. Calculated absorption spectra for the MAX phase compounds $Mo_2A_2AlC_3$ (A = Zr, Nb, Ta), respectively, for the [100] and [001] polarization directions are presented in Fig. 6(a). As these compounds are metallic, these material's show optical absorption from zero photon energy. The absorption coefficient for $Mo_2Zr_2AlC_3$, $Mo_2Nb_2AlC_3$, and $Mo_2Ta_2AlC_3$ rises with increasing photon energy to attain the highest values at 11.89, 8.84, 9.68 eV for [100] plane and 8.07, 8.96, 8.41 eV for [001] polarization direction within the *UV* region. Their pronounced absorption in the *UV* region highlights their suitability for applications that demand strong light absorption. The absorption coefficient steadily decreases with increasing photon energy. In almost the entire photon energy range, slight anisotropy is observed for both polarizations, which indicates a directional dependence in the material's optical properties.

Optical conductivity reflects a material's ability to conduct electricity when exposed to incident light [4]. Fig. 6(b) shows the calculated optical conductivity, $\sigma$. At low photon energy (~0.01 eV), all three compounds in both polarization directions show high conductivity, indicative of their metallic nature. As energy increases, optical conductivity decreases sharply, reaching low values by 23–30 eV. This indicates that these materials respond most strongly to low photon energy while showing reduced sensitivity to high photon energy. Between the [100] and [001] planes, the optical conductivity exhibits slight anisotropy across the photon energy range of 1–8 eV. The [100] plane consistently exhibits higher conductivity due to directional variations in bonding and electron density.

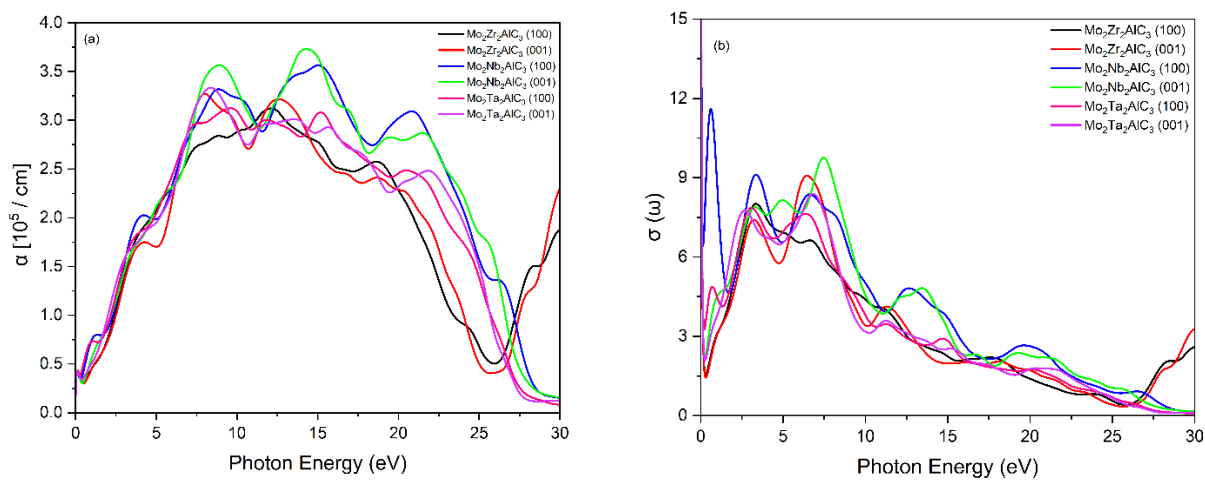

**Fig. 6:** (a) absorption coefficient α(ω) and (b) Optical conductivity.